\documentclass{aa}  
\usepackage{graphicx}
\usepackage{natbib}
\usepackage{txfonts}
\usepackage[debug]{hyperref}
\hypersetup{
    bookmarks=true,         % show bookmarks bar?
    unicode=false,          % non-Latin characters in Acrobat's bookmarks
    pdftoolbar=true,        % show Acrobat's toolbar?
    pdfmenubar=true,        % show Acrobat's menu?
    pdffitwindow=false,     % page fit to window when opened
    pdftitle={Beginning of activity in 67P/Churyumov-Gerasimenko and predictions for 2014/5},
    pdfauthor={Colin Snodgrass et al.},
    pdfsubject={},
    pdfkeywords={},         % list of keywords
    pdfnewwindow=true,      % links in new window
    colorlinks=true,        % false: boxed links; true: colored links
    linkcolor=gray,         % color of internal links
    citecolor=blue,         % color of links to bibliography
    filecolor=gray,         % color of file links
    urlcolor=gray           % color of external links
}

\usepackage{xcolor}
\definecolor{gray}{gray}{0.3}

\begin{document}
   \title{Beginning of activity in 67P/Churyumov-Gerasimenko and predictions for 2014/5
   \thanks{Based on observations
    collected at the European Southern Observatory, Paranal, Chile -
%    \href{http://archive.eso.org/wdb/wdb/eso/sched_rep_arc/query?progid=XX.C-XXXX}%
         {079.C-0687}
  \& 
%    \href{http://archive.eso.org/wdb/wdb/eso/sched_rep_arc/query?progid=XX.C-XXXX}%
         {080.C-0259}, and from archives.
}
}

   \author{C. Snodgrass
          \inst{1}
          \and
          C. Tubiana
          \inst{1}
          \and
          D. M. Bramich
          \inst{2}
          \and
          K. Meech
          \inst{3,4}
          \and
          H. Boehnhardt
          \inst{1}
          \and
          L. Barrera 
          \inst{5}
          }

   \institute{Max Planck Institute for Solar System Research, Max-Planck-Str. 2, 37191 Katlenburg-Lindau, Germany\\
              \email{snodgrass@mps.mpg.de}
         \and
             European Southern Observatory,
  Karl-Schwarzschild-Strasse 2,
  D-85748 Garching bei M\"{u}nchen, Germany
          \and
             Institute for Astronomy, 
             2680 Woodlawn Drive, Honolulu, HI 96822, USA%\\          
             \and
             University of Hawaii NASA Astrobiology Institute, Honolulu, HI 96822, USA
          \and
            Universidad Metropolitana de Ciencias de la Educacion (UMCE), Avda. Jose Pedro Alessandri 774, Santiago, Chile.
             }

   \date{Received ; accepted }

  \abstract
   {Comet 67P/Churyumov-Gerasimenko was selected in 2003 as the new target of the {\sl Rosetta} mission. It has since been the subject of a detailed campaign of observations to characterise its nucleus and activity.}
   {Here we present previously unpublished data taken around the start of activity of the comet in 2007/8, before its last perihelion passage. We constrain the time of the start of activity, and combine this with other data taken throughout the comet's orbit to make predictions for its likely behaviour during 2014/5 while {\sl Rosetta} is operating.}
   {A considerable difficulty in observing 67P during the past years has been its position against crowded fields towards the Galactic centre for much of the time. The 2007/8 data presented here were particularly difficult, and the comet will once again be badly placed for Earth-based observations in 2014/5. We make use of the difference image analysis (DIA) technique, which is commonly used in variable star and exoplanet research, to remove background sources and extract images of the comet. In addition, we reprocess a large quantity of archival images of 67P covering its full orbit, to produce a heliocentric lightcurve. By using consistent reduction, measurement and calibration techniques we generate a remarkably clean lightcurve, which can be used to measure a brightness--distance relationship and to predict the future brightness of the comet.}
   {We determine that the comet was active around November 2007, at a pre-perihelion distance from the Sun of 4.3 AU. The comet will reach this distance, and probably become active again, in March 2014. We find that the dust brightness can be well described by $Af\rho \propto r^{-3.2}$ pre-perihelion and $\propto r^{-3.4}$ post-perihelion, and that the comet has a higher dust-to-gas ratio than average, with log($Af\rho$/$Q$(H$_2$O)) = $-24.94\pm0.22$ cm s molecule$^{-1}$ at $r < 2$ AU. A model fit to the photometric data suggests that only a small fraction (1.4\%) of the surface is active.}
   {}


   \maketitle
%
%________________________________________________________________

\section{Introduction}

Comets are known for their unpredictable nature, and the process which drives their activity is still poorly understood. While the rise in ice sublimation due to increased heating from the Sun explains, in general terms, the increase in activity of a comet on the inbound leg of its orbit, the precise mechanism causing this activity is not known. The question of when and how a comet will become active is therefore difficult to answer, and there is a wide variation in activity levels at the same heliocentric distance for different comets. Understanding how activity starts and evolves as a comet approaches the Sun is one of the key goals of ESA's {\sl Rosetta} mission, which will rendezvous with comet 67P/Churyumov-Gerasimenko (hereafter 67P) in early 2014 and will follow it all the way to its next perihelion passage in 2015, and beyond. As part of the planning for this mission, and for a campaign of ground based observations in support of it, it is necessary to build up the best understanding possible of the behaviour of 67P before {\sl Rosetta}'s arrival.

Since 67P was selected as the {\sl Rosetta} target in 2003 there have been a number of studies of this comet. \citet{Schulz}, \citet{Lamy06} and \citet{Lara05} obtained the first dedicated characterisation observations after it was selected, when the comet was on the outbound leg of its orbit, and still highly active, following its perihelion passage in 2002. As the comet passed through aphelion its inactive nucleus was studied by \citet{Tubiana08,Tubiana11} and \citet{Lowry12}, before it was observed in an active state again as it returned to perihelion in 2008 \citep{Tozzi,Lara11}. Further authors have studied the longer lived dust trail associated with the comet \citep{Ishiguro08,Kelley08,Agarwal10}, the dust's polarisation \citep{Hadamcik10}, the nucleus thermal properties \citep{Kelley06,Lamy08,Kelley09}, and the morphology \citep{Vincent13} and composition \citep{Schleicher} of its coma over multiple apparitions.

What has been missing so far from the published literature is a reliable estimate of the point in the orbit that activity begins. In this paper we present data taken in 2007 and 2008 around the time of the expected onset, between the inactive nucleus observations taken in 2007 \citep{Tubiana11,Lowry12} and the active comet observations in 2008 \citep{Tozzi}. We use these to make estimates of the time of onset of activity in 2007/8, and from that make predictions for 2014.
 
To provide a broader context to the onset of activity observed in 2007/8, we also wish to study the activity of the comet throughout its orbit. The various studies of 67P in its active state that have been published in recent years have so far concentrated on a short segment of the orbit. \citet{Ferrin} published a heliocentric lightcurve based on mostly amateur observations, up to the 2002 apparition. To provide a consistent overview of the activity history of 67P, we downloaded and reduced data available in various professional observatory archives (primarily from ESO) and measured the comet brightness in various apertures. We use this to produce a heliocentric lightcurve based on consistent measurements, which is described in section \ref{hlc}, before looking at the implications for dust and gas production rates around the orbit in sections \ref{dust_production} and \ref{gas-section} respectively. We introduce a model that describes the activity in section \ref{karen-models}.

%__________________________________________________________________

\section{Observations and data reduction}

\subsection{VLT observations}

\begin{table}
\caption{Observational circumstances - 67P monitoring 2007/8.}
\label{obs-table}
\begin{center}
\begin{tabular}{rcccccc}
\hline
Date              	& r\tablefootmark{a}	&  $\Delta$\tablefootmark{b}	& $\alpha$\tablefootmark{c}		& Exp\tablefootmark{d}	& Filt.\\
%	& (AU)	& (AU) & (deg) & (s) & \\
\hline
%16.4.2007		& 4.95	& 4.52			& 11.0			& 10			& 60			& R		\\
%14.5.2007		& 4.86	& 4.06			& 8.1				& --			& --			& R		\\
%15.5.2007		& 4.85	& 4.05			& 7.9				& --			& --			& R		\\
%6.6.2007		& 4.78	& 3.80			& 3.7				& 10			& 60			& R		\\
%16.7.2007		& 4.63	& 3.72			& 6.0				& 143,3,3,7	& 100,420,100,140		& RBVI		\\
7/8/07		& 4.55	& 3.86			& 10.1			& 10	$\times$ 60	(4)		& R		\\
10/9/07		& 4.41	& 4.21			& 13.1			& 10	$\times$ 60	(5)		& R		\\
10/10/07	& 4.28	& 4.55			& 12.5			& 20	$\times$ 60	(10)		& R		\\
12/11/07	& 4.13	& 4.82			& 9.2				& 10	$\times$ 60	(7)		& R		\\
15/11/07	& 4.11	& 4.83			& 8.8				& 20	$\times$ 30	(0)		& R		\\
27/3/08		& 3.39	& 3.80			& 14.6			&  3$\times$15 + 4$\times$60 (6)	& VRI		\\
29/3/08		& 3.38	& 3.76			& 14.9			&  3$\times$15 + 4$\times$60	(6) & VRI		\\
30/3/08		& 3.38	& 3.74			& 15.0			&  3$\times$15 + 4$\times$60	(6) & VRI		\\
\hline
\end{tabular}
\end{center}
\tablefoottext{a}{Heliocentric distance (AU).}
\tablefoottext{b}{Geocentric distance (AU).}
\tablefoottext{c}{Phase angle (degrees).}
\tablefoottext{d}{Number of exposures $\times$ exposure time (seconds) taken. The number in parenthesis gives the number of frames used in the analysis.}
\end{table}

\begin{figure}
   \centering
   \begin{tabular}{@{} r @{} l @{}}
   \includegraphics[height=4.5cm]{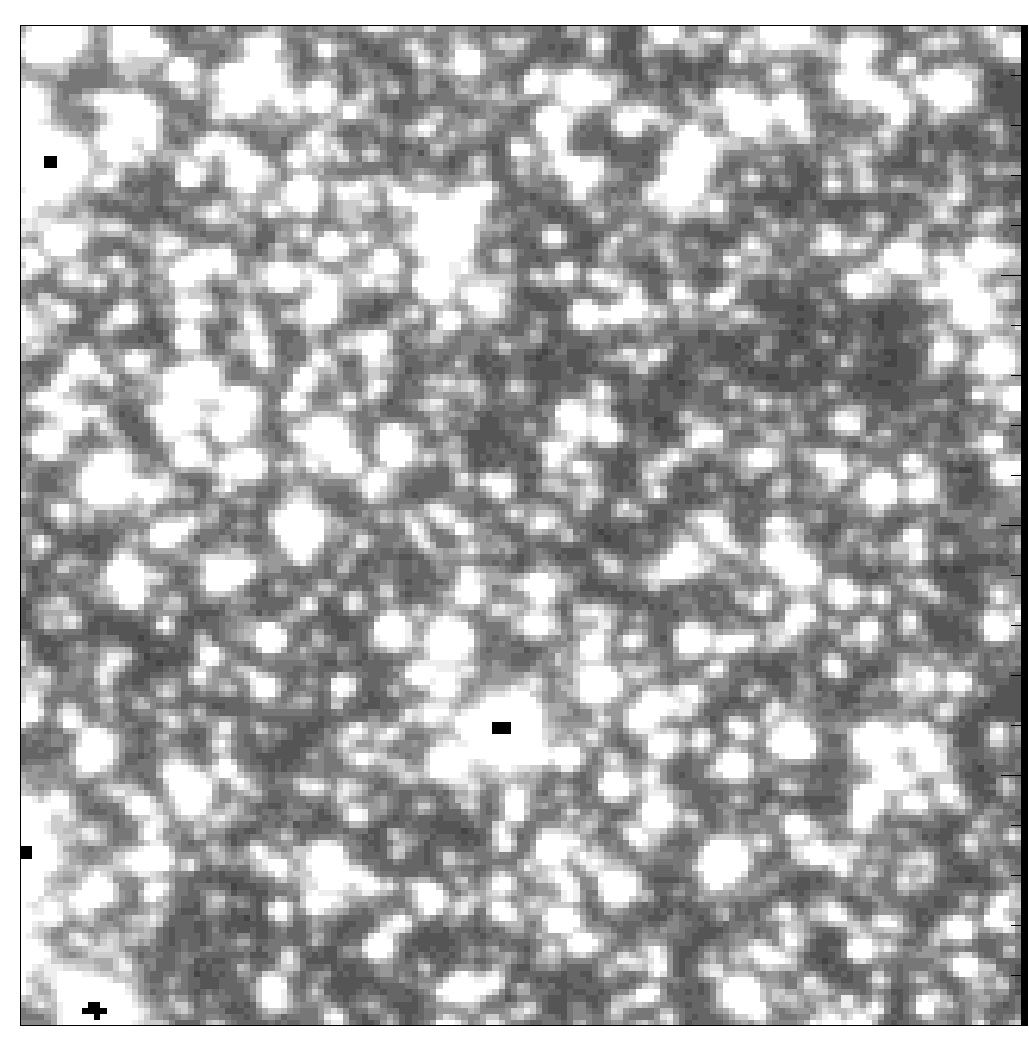} & 
   \includegraphics[height=4.5cm]{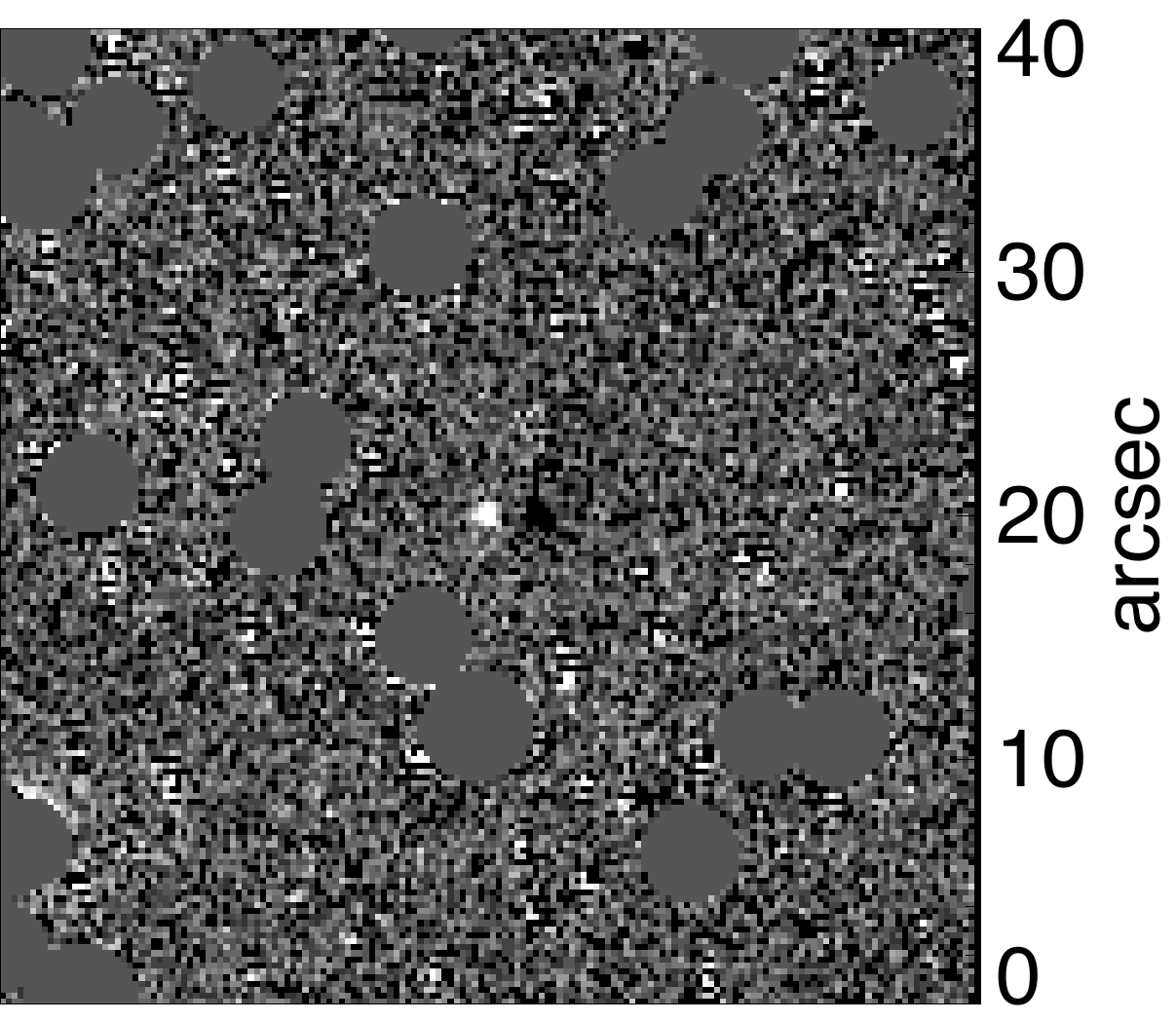} \\
   \end{tabular} 
   \caption{Example of the DIA method applied to 67P. This frame is taken from the October 2007 data set, and shows the original (left) and subtracted (right) versions of the same area, centred on the comet position. The comet is impossible to identify, much less perform photometry on, in such a crowded field, but is clearly revealed as a positive and negative pair in the subtracted frames. Saturated stars are masked in the difference image. North is up, East to the left.}
   \label{subtraction-image}
\end{figure}

Observations were obtained in service mode at the VLT using the FORS instrument in imaging mode. Images were mostly obtained in the $R$-band; details are given in table \ref{obs-table}. Standard reduction (bias level subtraction, flat fielding) was applied using IDL routines from the DanDIA / DanIDL packages\footnote{{\tt http://www.danidl.co.uk/}}. These packages provide an implementation of difference image analysis (DIA), which allows accurate subtraction of constant background sources in crowded fields, and includes corrections for changing seeing \citep{Bramich2008,Bramich13}. This technique has become a standard tool for variable star photometry in crowded fields, and especially microlensing, and we have recently demonstrated that it is also a powerful method for extracting moving (solar system) targets from dense background fields \citep{snodgrass+bramich}. Details on the method are given in these papers, but briefly the technique takes one `reference' frame (normally the one with the best seeing in a sequence) and applies a fitted kernel model to match this reference with each other frame. Subtracting these model frames from the data give images with all constant sources removed, revealing any variable stars or moving objects.   An example of the subtraction is shown in fig.~\ref{subtraction-image}. Use of these techniques was necessary for all data taken during 2007 due to the exceptionally crowded stellar fields, as the comet appeared near to the Galactic centre ($b \le 4^{\degr}$ throughout the period of observation). 

Once the background stars have been subtracted, standard aperture photometry can be used to measure the flux of the moving object. We follow the techniques that we have previously applied for comet nucleus photometry \citep[e.g.][]{Snodgrass05}, using an aperture with radius equal to the FWHM of the image point-spread-function (PSF) to maximise the signal-to-noise and an aperture correction to ensure that the full flux from the nucleus is included. When using DIA the PSF shape is modelled for each frame as part of the subtraction process, so this can be used to measure a high accuracy correction from a narrow aperture to the full PSF. This PSF shape was also compared with the radial profile of the comet in all frames to search for faint activity. Absolute flux calibration of the resulting photometry was performed based on observations of Landolt/Stetson standard star fields \citep{Landolt,Stetson} observed on the same nights, all of which were photometric. 

As these observations were not taken with DIA in mind, this data set presented some challenges. Firstly, there were a relatively large number of saturated stars in each frame, as even 60 seconds is a long exposure time for an 8m telescope when observing at low galactic latitude. Saturated stars cannot be subtracted and have to be masked. We were fortunate that the comet itself was not too close to saturated stars in most of the data, but the saturation still meant that care had to be taken in the convolution kernel calculation. Secondly, each image set covered only a short period of time ($\sim$40 minutes), with no separate background image (the star field without the comet) taken. This meant that the reference image needed to be taken from the comet frames, resulting in a negative image of the comet at the reference position in all frames, and the loss of 2-4 frames around the reference time where the comet flux could not be reliably measured due to this negative imprint. Enough well-subtracted frames remained to perform photometry on the comet for all nights except the 15th of November, when a combination of bad and highly variable seeing, high airmass, slow apparent motion of the comet, and nearby saturated stars meant that no useful comet images could be recovered. The total number of images used from each night is given in table \ref{obs-table}.

\subsection{Archival data}

\begin{table*}
\caption{Professional imaging of 67P, 1983-2010. Archival data from dates in bold face were used in this study.}

\small

\begin{center}
\begin{tabular}{r l l l c c c l}
\hline
UT Date & Instrument & PI & ESO Prog. ID. & r (AU) & $\Delta$ (AU) & $\alpha$ (deg) & Reference\\
\hline
12 Oct 1982 - 10 Mar 1983 & Lowell photometer & Schleicher & -- & 1.3 - 1.8 & 0.4 - 1.1 & 13 - 37 & \citet{Schleicher}\\
10-16  Apr 1991 & KPNO 2.1m+CCD & Mueller & -- & 4.9 & 3.9 & 4 & \citet{Mueller92}\\
\bf 25 Jun 1995 & \bf 2.2m+EFOSC & \bf Boehnhardt & \bf 055.F-0337  & \bf 2.5 & \bf 1.9 &  \bf 22 & \bf --\\
18 Nov 1995 - 14 Apr 1996 & Lowell photometer & Schleicher & -- & 1.3 - 1.6 & 1.0 - 1.8 & 34 - 48 & \citet{Schleicher}\\
\bf 12 Dec 1995 & \bf NTT+EMMI & \bf Boehnhardt & \bf 056.F-0133 & \bf 1.4 & \bf 1.0 & \bf 46 & \bf -- \\
09-11 Feb 1996 & OHP 1.93m + CARELEC & Weiler & -- & 1.3 & 1.2 & 46 & \citet{Weiler04}\\
14 Jun 1999 & JKT+CCD & Lowry & -- & 5.7 & 4.8 & 4 & \citet{Lowry03}\\
15 Jul 1999 & NTT+SUSI & Tancredi & 063.S-0504 & 5.7 & 5.1 & 8 & -- \\
25 May 2000 & UH 2.2m & Meech & -- & 5.3 & 4.3 & 4 & -- \\
30 Sep 2000 & UH 2.2m & Meech & -- & 5.0 & 5.1 & 8 & -- \\
\bf 17 Aug + 30 Sep 2001 & \bf UH 2.2m & \bf Meech & \bf -- & \bf 3.3 - 3.5 & \bf 2.8 - 3.0 & \bf 11 - 17 & \bf -- \\
\bf 09 Sep 2002 & \bf Kiso 1.05m Schmidt & \bf Ishiguro & \bf -- & \bf 1.3 & \bf 1.7 & \bf 36 & \bf \citet{Ishiguro08}\\
02 Dec 2002 + 01 Feb 2003 & Kiso 1.05m Schmidt & Ishiguro & -- & 1.8 - 2.2 & 1.4 - 1.6 & 18 - 33 & \citet{Ishiguro08}\\
\bf 11 Dec 2002 & \bf SDSS & \bf -- & \bf -- & \bf 1.8 & \bf 1.6 & \bf 32 & \bf -- \\
\bf 10 Feb 2003 & \bf 3.6m+EFOSC & \bf Boehnhardt & \bf 070.C-0505 & \bf 2.3 & \bf 1.4 & \bf 13 & \bf \citet{Schulz}\\
20 Feb + 27 Mar 2003 & TNG+DOLORES & Lara & -- & 2.4 - 2.6 & 1.4 - 1.7 & 9 - 10 & \citet{Lara05}\\
Feb - Apr 2003 & IAC-80+CCD & Lara & -- & 2.4 - 2.8 & 1.4 - 2.0 & 8 - 16 & \citet{Lara05}\\
07 Mar - 30 May 2003 & TLS 2m+CCD & Weiler & -- & 2.5 - 3.1 & 1.5 - 2.8 & 4 - 19 & \citet{Weiler04}\\
\bf 08 Mar 2003 & \bf 3.6m+EFOSC & \bf Boehnhardt & \bf 070.C-0505 & \bf 2.5 & \bf 1.5 & \bf 4 & \bf \citet{Schulz}\\
12 - 13 Mar 2003 & HST+WFPC2 & Lamy & -- & 2.5 & 1.5 & 5 & \citet{Lamy06}\\
\bf 25 Mar 2003 & \bf NTT+EMMI & \bf LSO tech & \bf -- & \bf 2.6 & \bf 1.7 & \bf 9 & \bf -- \\
\bf 30 Apr - 24 Jun 2003 & \bf VLT+FORS & \bf Schulz & \bf 270.C-5035 & \bf 2.9 - 3.2 & \bf 2.3 - 3.2 & \bf 18 - 19 & \bf \citet{Schulz}\\
29+30 May 2003 & UH 2.2m & Meech & -- & 3.0 & 2.8 & 19 & -- \\
25-27 Jun 2003 & Hale 200" + LFC & Kelley & -- & 3.2 & 3.4 & 18 & \citet{Kelley08}\\
\bf 22-26 Feb 2004 & \bf NTT+EMMI/SUSI & \bf LSO tech & \bf -- & \bf 4.5 & \bf 4.0 & \bf 12 & \bf \citet{Lowry12}\\
14 Apr 2004 & VLT+FORS & Boehnhardt & 073.C-0346 & 4.7 & 3.7 & 2 & \citet{Tubiana11}\\
18-21 Apr 2004 & 2.2m+WFI & Agarwal & 072.A-9011 & 4.7 & 3.7 & 1 & \citet{Agarwal10}\\
23+27 Apr 2004 & NTT+EMMI & Lowry & 073.C-0061 & 4.7 & 3.7 & 1 & -- \\
\bf 30 Apr + 01 May 2004 & \bf 3.6m+EFOSC & \bf Stuewe & \bf 073.C-0846 & \bf 4.7 & \bf 3.7 & \bf 2 & \bf -- \\
13 Jun 2004 & 3.6m+EFOSC & Rauer & 073.C-0571 & 4.9 & 4.2 & 10 & -- \\
\bf 16 Jun 2004 & \bf VLT+FORS & \bf Boehnhardt & \bf 073.C-0346 & \bf 4.9 & \bf 4.3 & \bf 10 & \bf \citet{Tubiana08}\\
21+22 Jun 2004 & UH 2.2m & Meech & -- & 4.9 & 4.4 & 11 & -- \\
10-14 May 2005 & NTT+EMMI & Lowry & 075.C-0247 & 5.6 & 4.6 & 0.1 - 0.8 & \citet{Lowry12}\\
11 May - 01 Jun 2005 & CFHT+Megacam & Ishiguro & -- & 5.6 & 4.6 & 0.1 - 3 & -- \\
04-05 Aug 2005 & VLT+VIMOS & Gruen & 275.B-5034 & 5.7 & 5.5 & 10 & -- \\
\bf 25 May - 01 Jun 2006 & \bf VLT+FORS & \bf Barrera & \bf 077.C-0609 & \bf 5.6 & \bf 4.6 & \bf 1 & \bf \citet{Tubiana08}\\
\bf 17-22 Aug 2006 & \bf VLT+FORS & \bf Boehnhardt & \bf 277.C-5038 & \bf 5.5 & \bf 5.3 & \bf 10 & \bf \citet{Tubiana08}\\
16 Apr - 16 Jul 2007 & VLT+FORS & Barrera & 079.C-0687 & 4.6 - 5.0 & 3.7 - 4.5 & 4 - 11 & \citet{Tubiana11}\\
16-22 Jul 2007 & NTT+EMMI & Lowry & 079.C-0384 & 4.6 & 3.7 & 6 & \citet{Lowry12}\\
\bf 07 Aug 2007 & \bf VLT+FORS & \bf Barrera & \bf 079.C-0687 & \bf 4.6 & \bf 3.8 & \bf 10 & \bf this work \\
\bf 10 Sep 2007 & \bf VLT+FORS & \bf Barrera & \bf 079.C-0687 & \bf 4.4 & \bf 4.2 & \bf 13 & \bf this work \\
14 Sep 2007 & NTT+EMMI & Lowry & 079.C-0384 & 4.4 & 4.2 & 13 & \citet{Lowry12}\\
\bf 10 Oct 2007 & \bf VLT+FORS & \bf Barrera & \bf 080.C-0259 & \bf 4.3 & \bf 4.5 & \bf 13 & \bf this work \\
\bf 12-15 Nov 2007 & \bf VLT+FORS & \bf Barrera & \bf 080.C-0259 & \bf 4.3 & \bf 4.6 & \bf 12 & \bf this work \\
\bf 27-30 Mar 2008 & \bf VLT+FORS & \bf Barrera & \bf 080.C-0259 & \bf 3.4 & \bf 3.8 & \bf 15 & \bf this work \\
30-31 May 2008 & VLT+FORS & Schulz & 281.C-5004 & 3.0 & 2.5 & 19 & -- \\
\bf 01-05 Jun 2008 & {\bf VLT+FORS}/ISAAC & \bf Tozzi & \bf 381.C-0123 & \bf  3.0 & \bf 2.5 & \bf 19 & \bf \citet{Tozzi}\\
\bf 02-04 Jul 2008 & \bf VLT+FORS & \bf Schulz & \bf 281.C-5004 & \bf 2.8 & \bf 1.9 & \bf 14 & \bf -- \\
06-10 Jul 2008 & 2.2m+WFI & Gruen & 081.A-9019 & 2.7 & 1.9 & 14 & \citet{Agarwal09}\\
01-16 Aug 2008 & 2.2m+WFI & LSO tech & -- & 2.4 - 2.6 & 1.5 - 1.6 & 6 & -- \\
\bf 10 Aug 2008 & \bf VLT+FORS & \bf Schulz & \bf 281.C-5004 & \bf 2.5 & \bf 1.5 & \bf 5 & \bf -- \\
\bf 03-07 Sep 2008 & {\bf VLT+FORS}/ISAAC & \bf Tozzi & \bf 381.C-0123 & \bf 2.3 & \bf 1.4 & \bf 15 & \bf \citet{Tozzi}\\
\bf 13 Sep 2008 & \bf VLT+FORS & \bf Schulz & \bf 281.C-5004 & \bf 2.2 & \bf 1.4 & \bf 19 & \bf -- \\
\bf 21-29 Oct 2008 & {\bf VLT+FORS}/ISAAC & \bf Tozzi & \bf 082.C-0740 & \bf 1.9 & \bf 1.5 & \bf 31 & \bf \citet{Tozzi}\\
11 Dec 2008 & IAA 1.52m + CCD & -- & -- & 1.6 & 1.7 & 35 & \citet{Fulle10}\\
25-27 Dec 2008 & IGO 2m & -- & -- & 1.5 & 1.7 & 36 & \citet{Hadamcik10} \\
13 Jan 2009 & TNG+DOLORES & Tozzi & -- & 1.4 & 1.7 & 36 & \citet{Tozzi}\\
\bf 25 Jan - 12 Mar 2009 & \bf Lulin 1m + CCD & \bf Lara & \bf -- & \bf 1.2 - 1.3 & \bf 1.7 & \bf 36 & \bf \citet{Lara11}\\
\bf 28 Jan 2009 & \bf NTT+EFOSC & \bf Lowry & \bf 082.C-0517 & \bf 1.3 & \bf 1.7 & \bf 36 & \bf  -- \\
\bf 19 Mar 2009 & \bf CA 2.2m+CAFOS & \bf Lara & \bf -- & \bf 1.3 & \bf 1.7 & \bf 35 & \bf  \citet{Lara11}\\
17-19 Mar 2009 & OHP 0.8m & -- & -- & 1.3 & 1.7 & 35 & \citet{Hadamcik10}\\
30 Apr - 01 May 2009 & IGO 2m & -- & -- & 1.4 & 2.0 & 29 & \citet{Hadamcik10} \\
01 Feb - 29 Mar 2010 & VLT+FORS & Boehnhardt & 384.C-0115 & 3.4 - 3.7 & 2.7 - 3.0 & 3 - 16 & -- \\
\bf 14-16 Feb 2010 & \bf NTT+EFOSC & \bf Snodgrass & \bf 084.C-0594 & \bf 3.5 & \bf 2.9 & \bf 14 & \bf -- \\
16 Mar 2010 & SOAR+SOI & Barrera & -- & 3.6 & 2.7 & 8 & -- \\
09 Apr 2010 & SOAR+SOI & Barrera & -- & 3.8 & 2.8 & 1 & -- \\
\bf 14 Jul + 30 Aug 2010 & \bf NTT+EFOSC & \bf Lowry & \bf 185.C-1033 & \bf 4.3 - 4.5 & \bf 4.3 - 5.1 & \bf 9 - 14 & \bf -- \\
\hline
\end{tabular}
\end{center}
\label{archive-obs}
\end{table*}%

\begin{figure}
   \centering
   \includegraphics[width=\columnwidth]{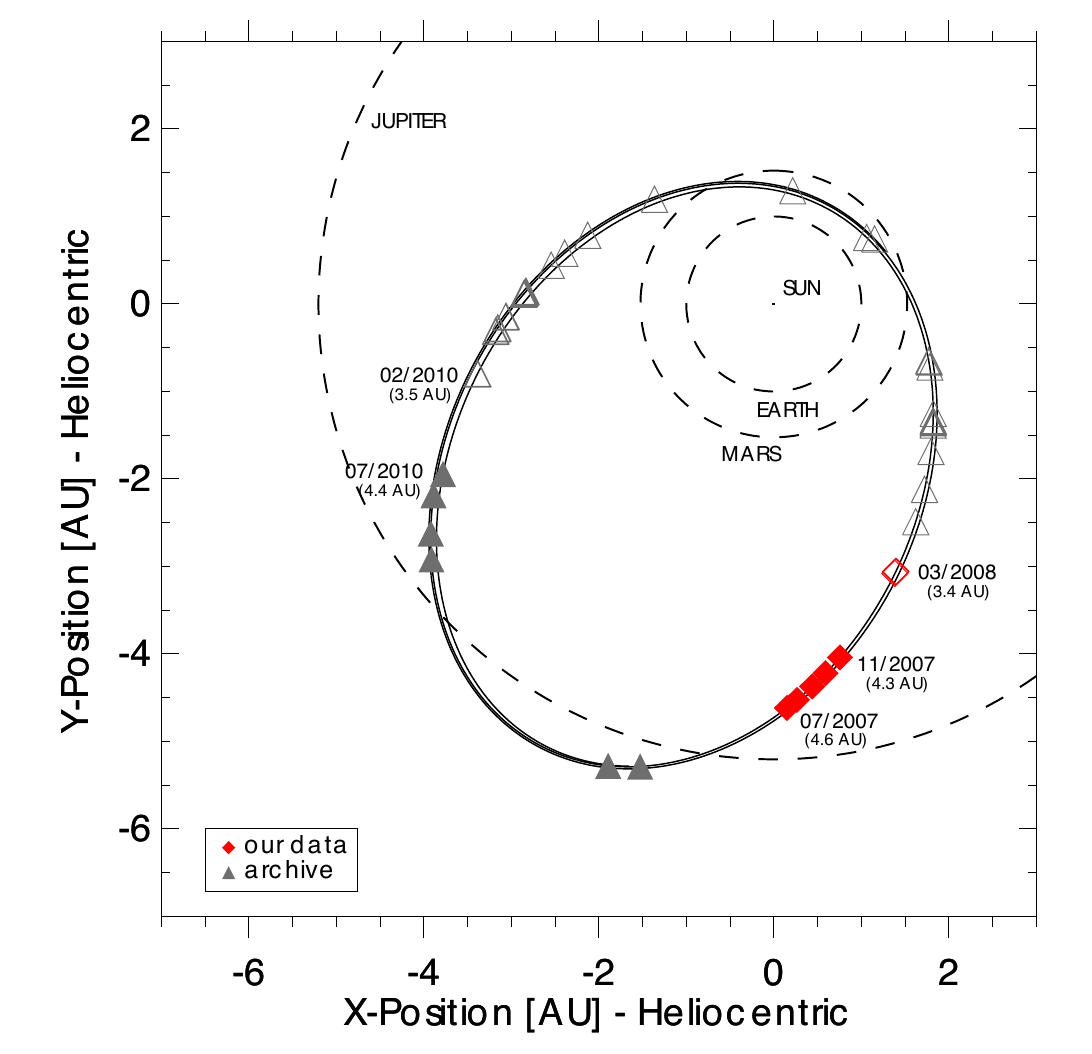} 
   \caption{Position of 67P in its orbit at the time of the observations used in this paper. Open symbols indicate visible activity, closed symbols indicate an apparently inactive nucleus. The VLT data presented in detail are marked by red diamonds, while the archival data are marked by triangles. The comet moves in an counter-clockwise fashion in this plot.}
   \label{orbit-plot}
\end{figure}

There is a large set of archival data on 67P, with professional observations on approximately 200 individual nights to date (mostly snap-shots, but some full nights of lightcurve observations). Table \ref{archive-obs} summarises all of these observations, and indicates the ones used in this study. We selected only $R$-band images, to give an overall description of the brightness of the comet due to reflected sunlight, as this wavelength range is relatively free of gas emission lines. Figure \ref{orbit-plot} shows the position of the comet in its orbit at the date of each selected observation.

The vast majority of the data (especially since the selection of the comet as the new {\sl Rosetta} target in 2003) has been taken at ESO, and is therefore available through the ESO archival service\footnote{{\tt http://archive.eso.org}}. These data were largely taken by a few groups (led by Schulz, B{\"o}hnhardt, Barrera, Tubiana, Lowry and Tozzi) and have mostly been published elsewhere already. However, each group reported measurements in different ways, so we reprocessed all the data and performed the measurements in a consistent manner to give a uniform data set. Standard reduction was performed using DanDIA tools, although no image subtraction was necessary for the frames used. To allow automated processing of such a large data set, master bias and flat field frames covering long periods ($\sim 1$ year) were used for each instrument. While these introduce some systematic errors, this is not a problem at the level of photometric accuracy ($\sim 0.1$ mag) necessary for studying the total brightness evolution of the comet around its orbit. It is a testament to the stability of the ESO instruments that the reductions performed in this way are actually very good. 

Calibration of each frame was performed, where possible, using field stars of known brightness and colour from photometric surveys. Throughout most of 2003 and some of 2010 the comet was in an area of sky covered by the Sloan Digital Sky Survey (SDSS), so calibration was performed using field stars in this catalogue. 
The 1995 data were calibrated using field stars available in the AAVSO Photometric All-Sky Survey (APASS) catalogue, which covers the full sky but unfortunately could not be used for most of the data due to either the small field of view or the long exposure time used, and consequently a lack of non-saturated catalogue stars. Where catalogue stars were not available in the comet field we determined nightly zeropoints using Landolt/Stetson star fields observed on the same night (assuming default extinction and colour term values to allow quick automated fitting). In a few cases where this was not possible we used the tabulated nightly zeropoints available for VLT instruments on the ESO web page, or default instrumental values. Where possible APASS and VLT nightly zeropoint calibrations were also checked against the SDSS; we find that the APASS and SDSS photometry give consistent results, while there is an offset of up to 0.5 mag between the VLT tabulated / default zeropoints and SDSS in 2003 - the points calibrated this way are treated with caution. The calibration source for each night is listed in table \ref{archive-results}. 

In addition to the ESO data, we also took some frames from the Japanese SMOKA archive (observations with the Kiso 1.05m Schmidt) and the SDSS (which serendipitously observed the comet in December 2002). Kiso observations were reduced in the same way as the ESO data, and calibrated using the APASS \& SDSS catalogues, while the SDSS data is already reduced and photometrically calibrated into the SDSS system. For the SDSS image, new aperture photometry was measured on the comet, and the SDSS $r$-band magnitudes then converted to Cousins (Landolt) $R$-band using\footnote{{\tt http://www.sdss3.org/dr9/algorithms/sdssUBVRITransform.php}}
\begin{equation}
R = r - 0.1837(g - r) - 0.0971
\end{equation}
and solar ($g-r$) = 0.45 \citep{Holmberg06}. Observations are also available in the HST and CFHT (Megacam) archives, although we chose not to include these for simplicity, as the dates of these observations are near to dates covered by ESO observations. Finally, images from the University of Hawaii 2.2m telescope were identified in Karen Meech's archive of comet observations, although these were unfortunately of limited use as the comet was either not detected or merged with background stars in many of the frames (these snap-shot images are not suitable for DIA processing, as no image of the field without the comet is available). The solar system object search tool of the Canadian Astronomy Data Centre\footnote{{\tt http://www1.cadc-ccda.hia-iha.nrc-cnrc.gc.ca/ssos/}} \citep{SSOS} proved exceptionally useful in identifying archival observations in addition to those we already knew of.

%__________________________________________________________________

\section{2007/8 results}\label{activity_start}

\subsection{Beginning of activity}

\begin{figure}
   \centering
   \includegraphics[width=\columnwidth]{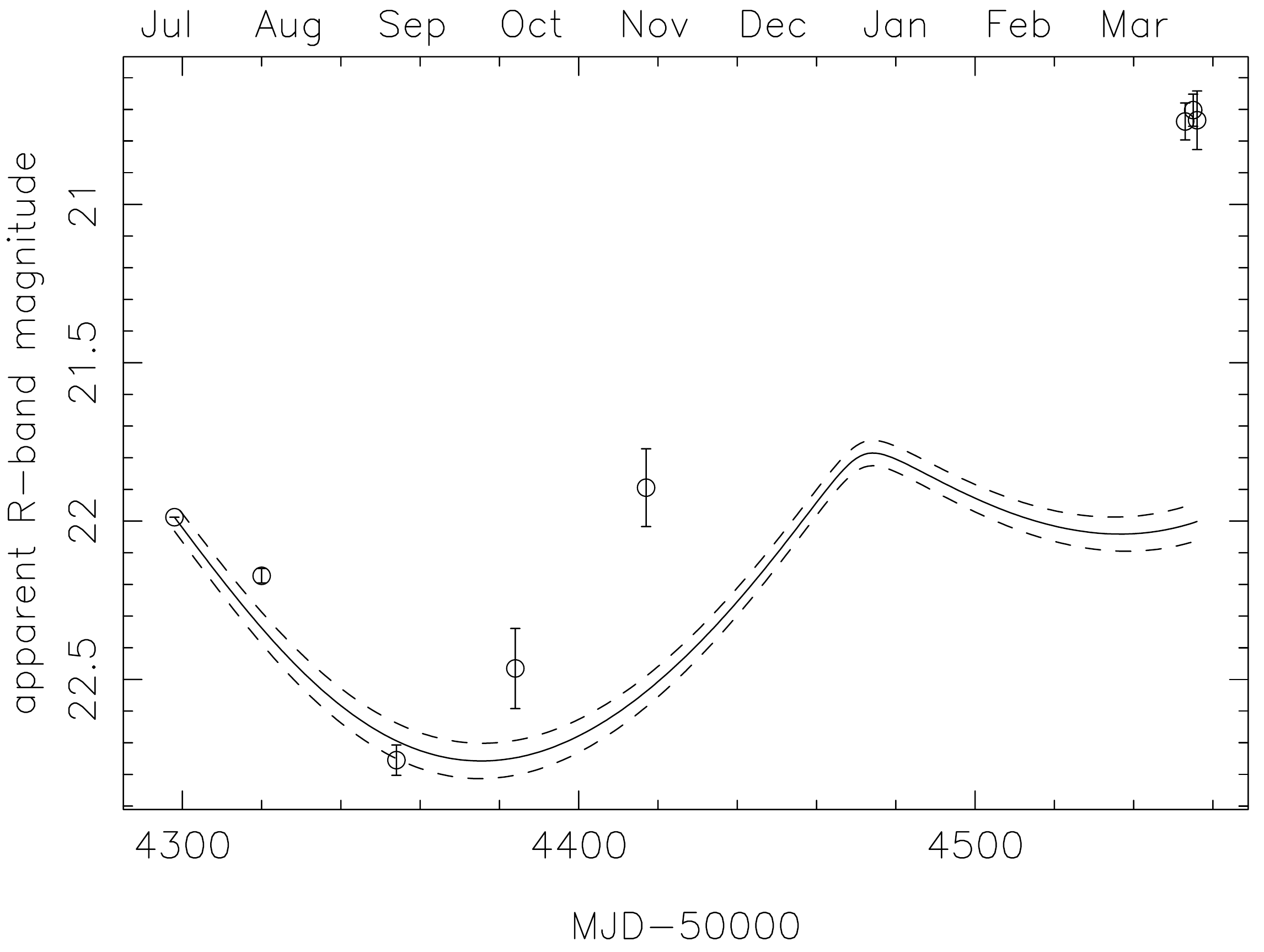} 
   \caption{Photometry covering the period July 2007 to March 2008, showing the apparent $R$-band magnitude of the comet (corrected for rotational phase). The solid line shows the predicted magnitude of the bare nucleus (based on the July 2007 data; \citet{Tubiana11}), with the dashed lines showing the 1$\sigma$ uncertainty on this. The comet appears to be significantly brighter than expected by November 2007, indicating some unresolved activity within the aperture at this point (at a heliocentric distance of 4.3 AU). By March 2008, when the comet is visibly active in images, the difference is more than 1 magnitude.}
   \label{turn-on-phot}
\end{figure}

We plot the measured $R$-band magnitude of the comet during 2007/8 in fig.~\ref{turn-on-phot}, which also shows the predicted nuclear magnitude based on the observations of the inactive nucleus in July 2007 and measurement of the nucleus phase function by \citet{Tubiana08,Tubiana11}. Our VLT observations were short sequences (covering only a small fraction on the comet's $\sim 12.7$ hour rotation period), and we are looking for small changes in brightness (less than the $\Delta m \approx 0.4$ mag.~variation due to the nucleus lightcurve). We therefore applied a correction to the photometry, by using the nucleus shape and pole model published by \cite{Lowry12} to calculate the predicted rotational lightcurve phase at the time of the observations, and the corresponding offset from the mean effective magnitude, as would be observed for a non-varying (spherical) nucleus. As \citet{Lowry12} also use July 2007 data in their model, the uncertainty in the extrapolation of rotational phase is small ($\sim 0.1\%$ by November 2007). The corrections to the photometry due to rotation phase are at most 0.13 magnitudes (for the August 2007 data), and smaller than the uncertainties for all other months (Sep: -0.01 mag.; Oct:+0.01; Nov: +0.03; Mar: n/a). For each month we plot the average magnitude from the individual measurements, with the error bar given by the standard deviation on this mean. The individual photometric measurements, without correction for rotational phase, are given in table \ref{VLT-phot}.

\begin{table}
\caption{Photometry from the 2007/8 VLT data (without rotational phase correction).}
\label{VLT-phot}
\begin{center}
\begin{tabular}{c c c c}
\hline
Date & MJD              	& $m_R$	& $\sigma_R$\\
\hline
7/8/2007 & 54320.05823 & 22.32 & 0.04 \\
& 54320.05951 & 22.27 & 0.03 \\
& 54320.06068 & 22.31 & 0.03 \\
& 54320.06194 & 22.31 & 0.04 \\
10/9/2007 & 54354.03934 & 22.76 & 0.06 \\
& 54354.04065 & 22.74 & 0.05 \\
& 54354.04185 & 22.75 & 0.06 \\
& 54354.04317 & 22.80 & 0.06 \\
& 54354.04437 & 22.67 & 0.05 \\
10/10/2007 & 54384.01135 & 22.40 & 0.06 \\
& 54384.01252 & 22.29 & 0.06 \\
& 54384.01358 & 22.32 & 0.06 \\
& 54384.01476 & 22.54 & 0.07 \\
& 54384.01583 & 22.60 & 0.08 \\
& 54384.01971 & 22.50 & 0.07 \\
& 54384.02079 & 22.70 & 0.08 \\
& 54384.02303 & 22.50 & 0.07 \\
& 54384.02420 & 22.38 & 0.06 \\
& 54384.02527 & 22.51 & 0.07 \\
12/11/2007 & 54417.00344 & 21.80 & 0.14 \\
& 54417.00449 & 21.87 & 0.12 \\
& 54417.00560 & 22.01 & 0.13 \\
& 54417.00664 & 21.86 & 0.12 \\
& 54417.00787 & 21.83 & 0.10 \\
& 54417.00901 & 21.95 & 0.12 \\
& 54417.01177 & 22.15 & 0.13 \\
27/3/2008 & 54553.37439 & 20.67 & 0.04 \\
& 54553.37554 & 20.82 & 0.04 \\
& 54553.37674 & 20.68 & 0.02 \\
& 54553.37785 & 20.77 & 0.02 \\
& 54553.37918 & 20.72 & 0.02 \\
& 54553.38026 & 20.77 & 0.02 \\
29/3/2008 & 54555.38642 & 20.80 & 0.04 \\
& 54555.38699 & 20.71 & 0.03 \\
& 54555.38801 & 20.69 & 0.02 \\
& 54555.38920 & 20.68 & 0.02 \\
& 54555.39027 & 20.67 & 0.02 \\
& 54555.39145 & 20.66 & 0.02 \\
30/3/2008 & 54556.38324 & 20.90 & 0.04 \\
& 54556.38381 & 20.78 & 0.04 \\
& 54556.38534 & 20.66 & 0.02 \\
& 54556.38642 & 20.67 & 0.02 \\
& 54556.38750 & 20.68 & 0.02 \\
& 54556.38858 & 20.72 & 0.02 \\

\hline
\end{tabular}
\end{center}
\end{table}

\begin{figure}
   \centering
   \includegraphics[width=\columnwidth]{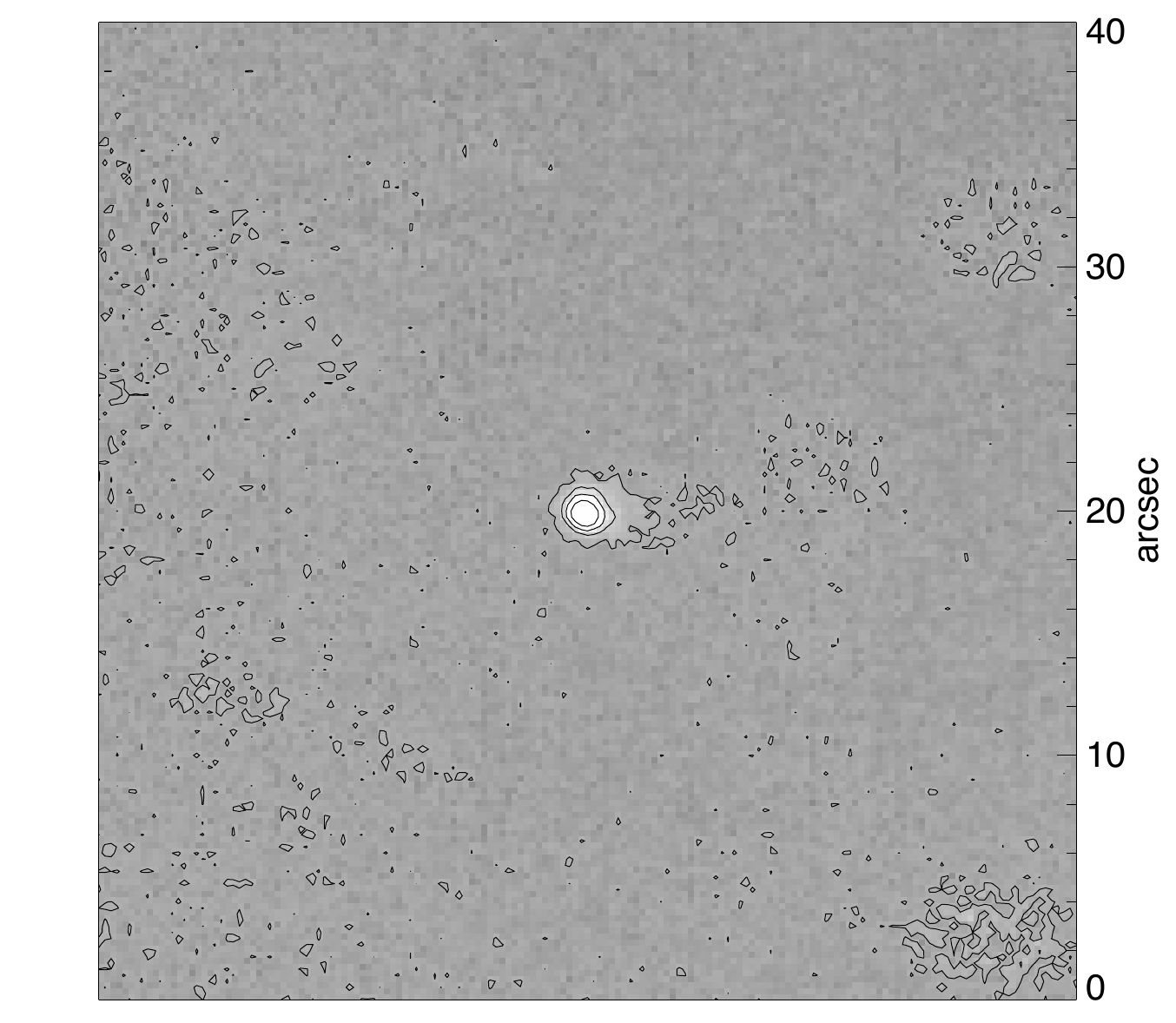} 
   \caption{Median $R$-band image from 12$\times$60 second images over 3 nights in March 2008. By this point (at $r=3.4$ AU) the comet is visibly active, with a faint tail to the West (right in this image; North is up).}
   \label{March08img}
\end{figure}

The offset from the predicted nucleus magnitude is clear for the March 2008 data, where faint activity is also apparent in the images (fig. \ref{March08img}). We can therefore be certain that the comet was active in March 2008, at 3.4 AU from the Sun on its inbound leg. The photometry also shows a significant amount of extra flux in November 2007 (MJD 54417), suggesting some weak activity at that time, when the comet was at 4.3 AU from the Sun. The October data also suggest a possible excess flux, although the uncertainty on this data point makes it consistent with the inactive nucleus at a 2$\sigma$ level, while the September data set (where the comet was detected at high signal-to-noise, well away from residuals due to saturated stars) indicates that activity had not yet reached a detectable level at 4.4 AU. 

\begin{figure}
   \centering
   \begin{tabular}{l  l}
   \includegraphics[width=0.7\columnwidth]{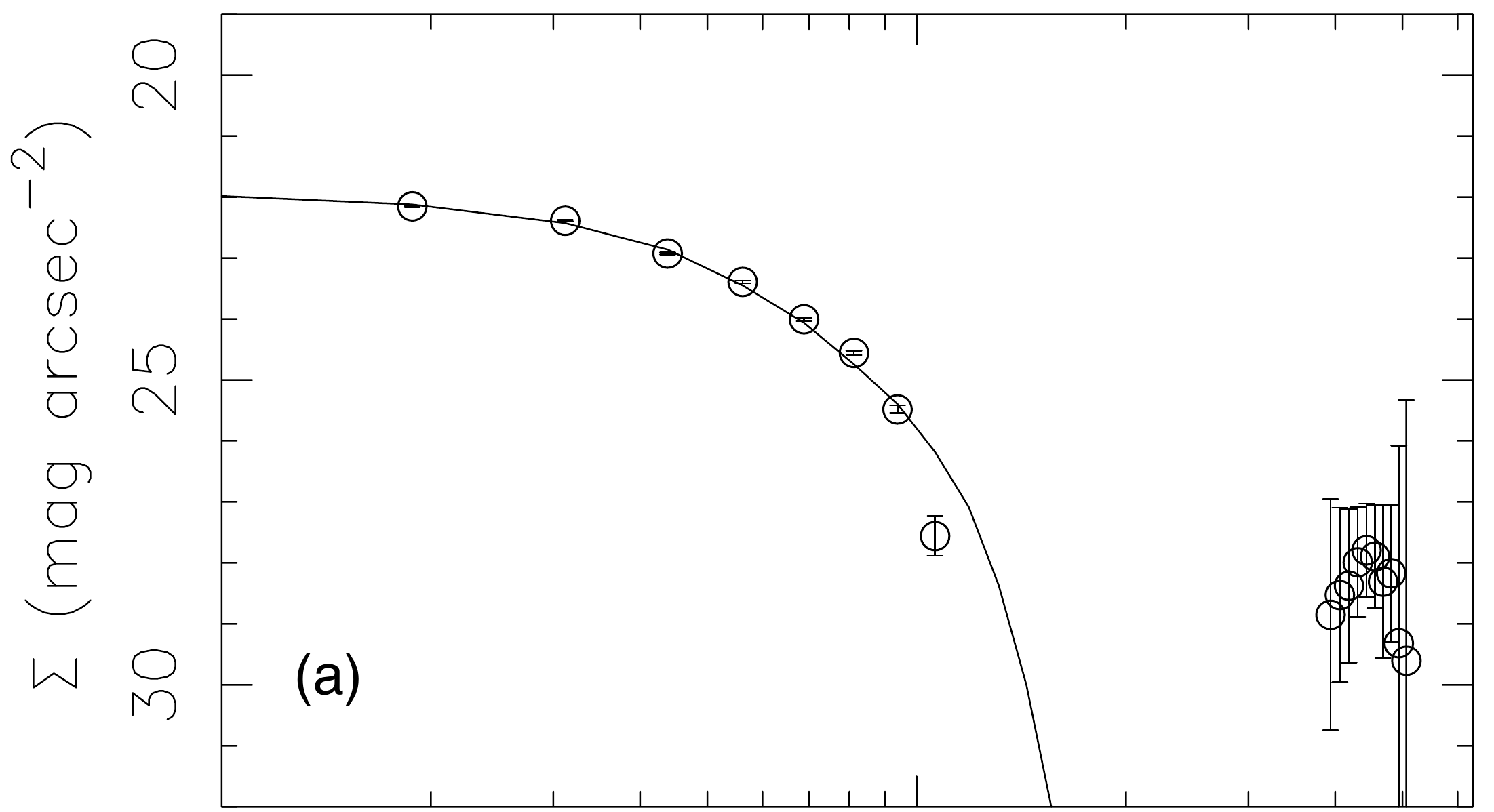}  & \raisebox{3cm}{(a)} \\
   \includegraphics[width=0.7\columnwidth]{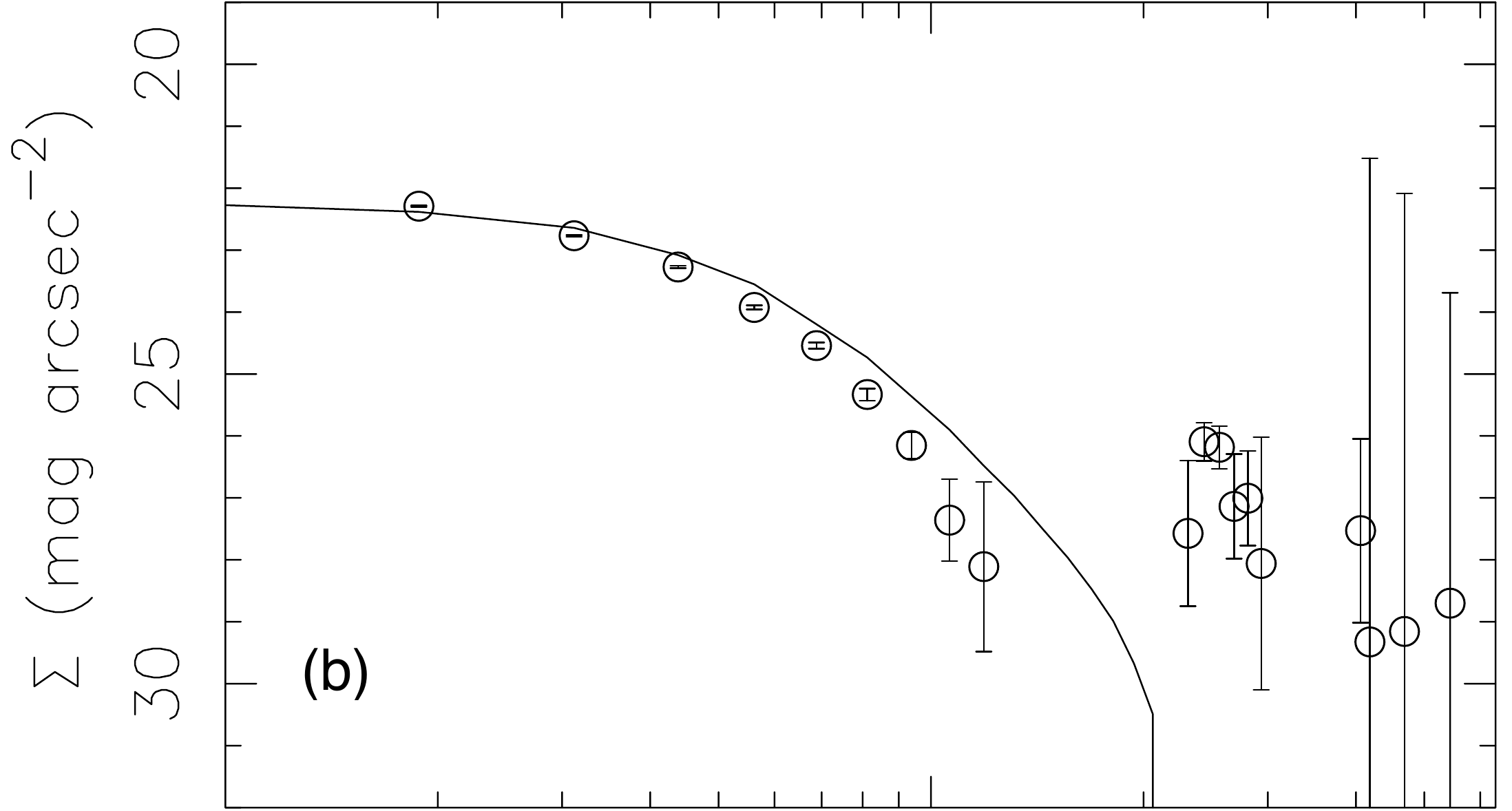}  & \raisebox{3cm}{(b)} \\
   \includegraphics[width=0.7\columnwidth]{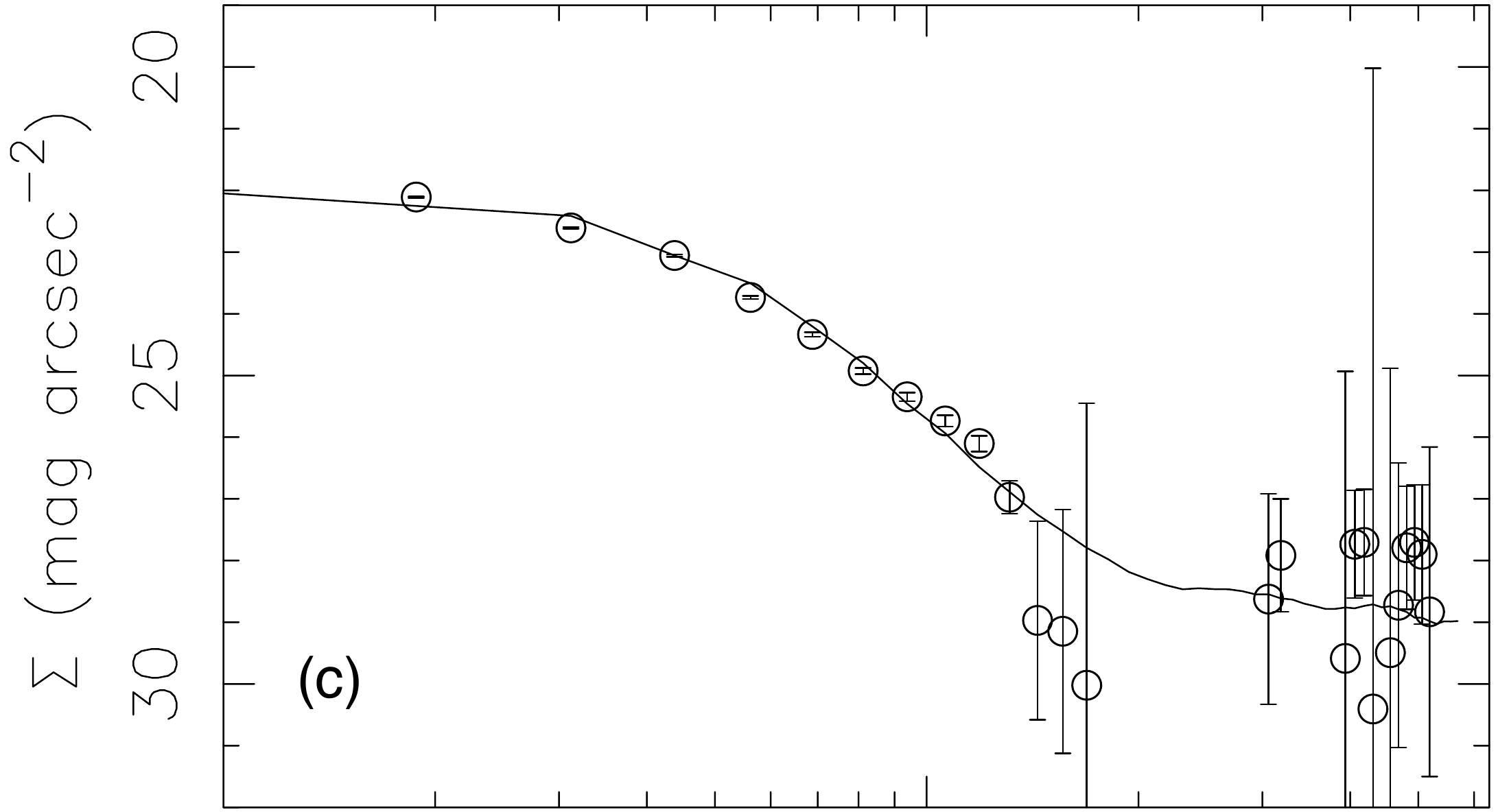}  & \raisebox{3cm}{(c)} \\
   \includegraphics[width=0.7\columnwidth]{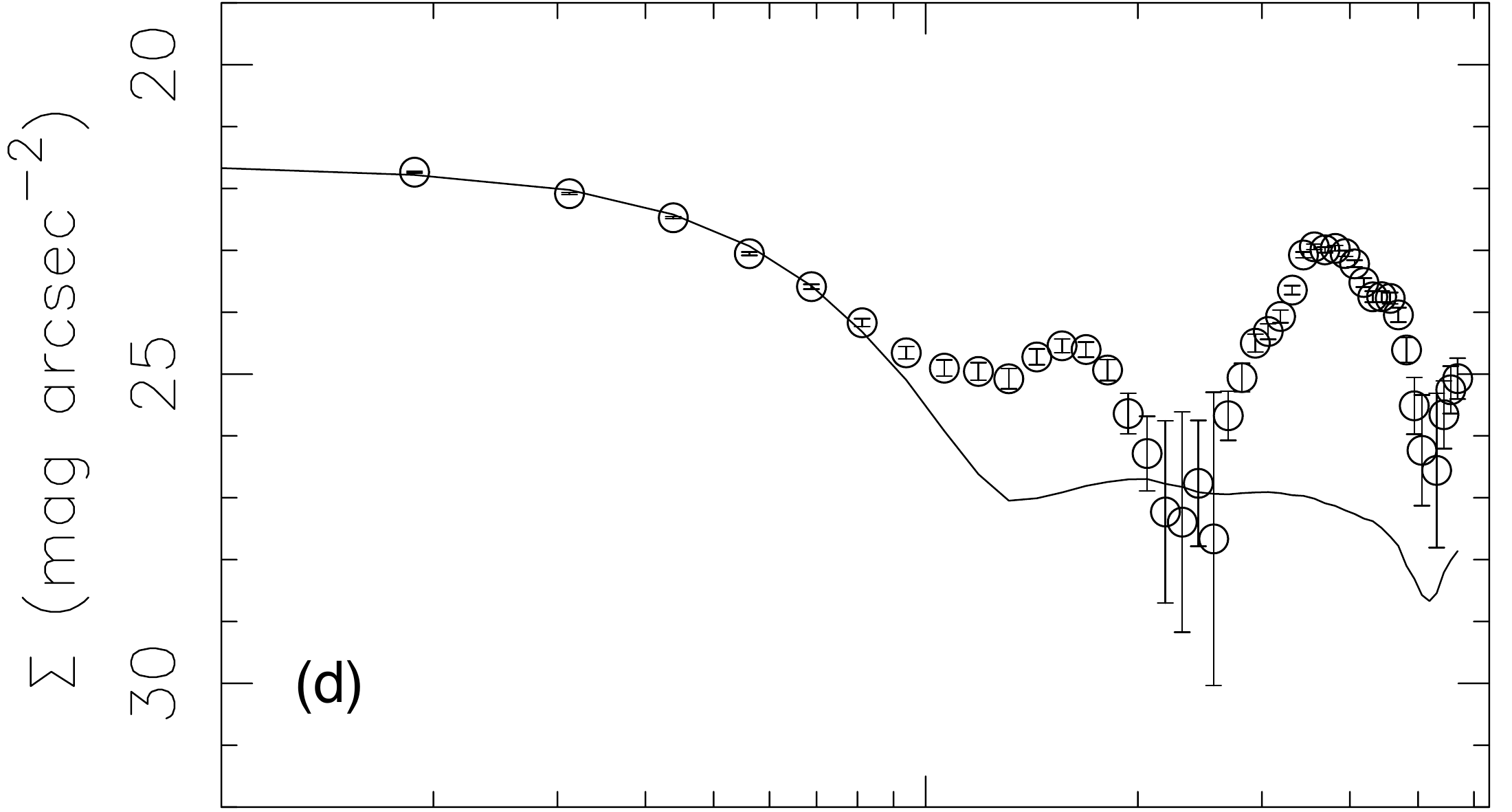} & \raisebox{3cm}{(d)} \\
   \includegraphics[width=0.7\columnwidth]{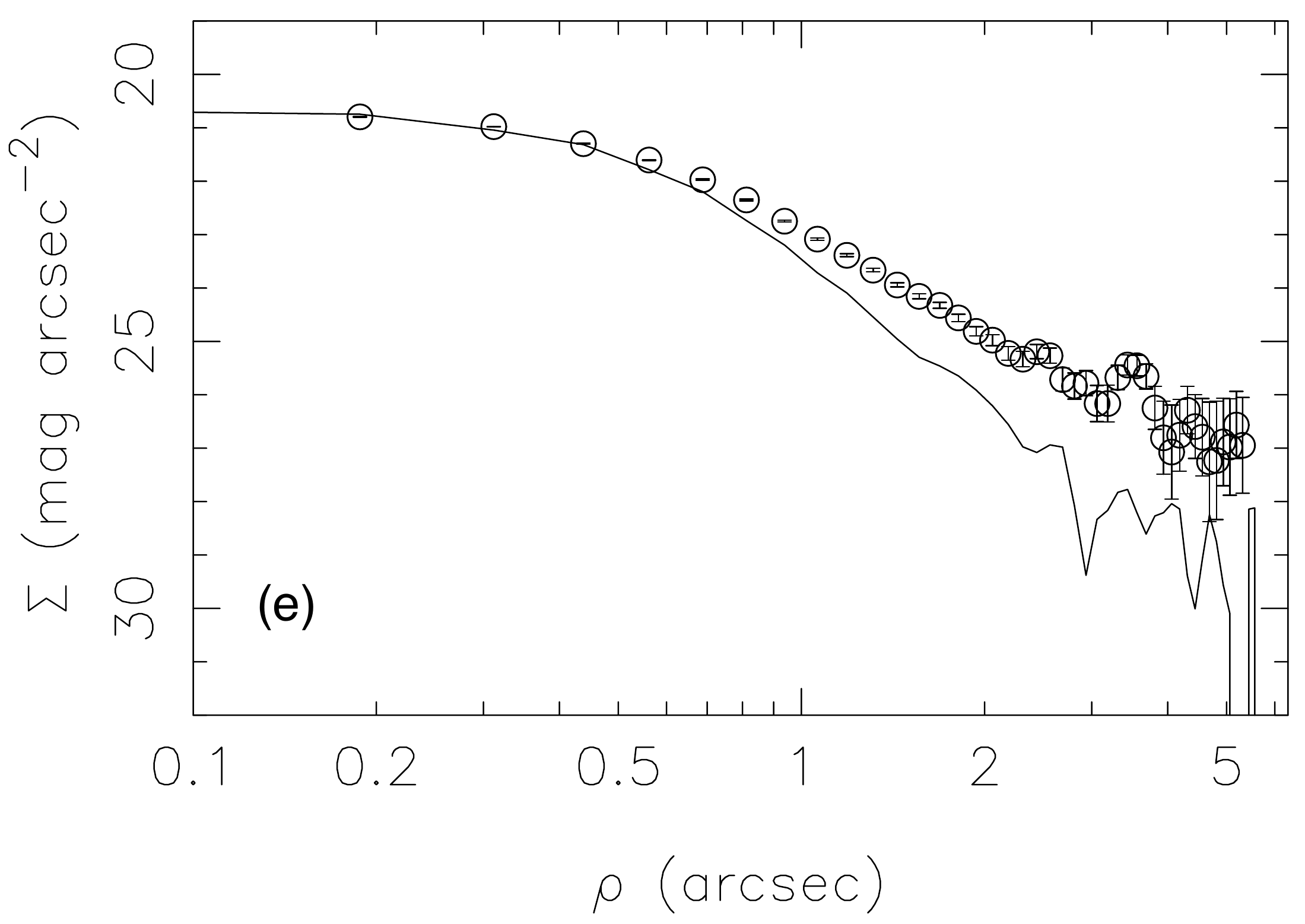} & \raisebox{4cm}{(e)} \\
  \end{tabular}   
   \caption{Surface brightness profiles for the comet in (a) August, (b) September, (c) October, (d) November 2007, and (e) March 2008. The points show the comet profile, while the line shows the image PSF. (a) - (c) show star-like profiles, while (e) shows clear activity. (d) is difficult to interpret due to residuals in the image from saturated stars (the bumps beyond $\rho=1\arcsec$). The gaps in the comet profiles in (a) - (c) around $\rho \sim 2\arcsec{}$ are due to the negative imprint from the reference image.}
   \label{sbps}
\end{figure}

In addition to measuring the brightness of the comet, we also test for activity by comparing its surface brightness profile with the shape of the PSF. Profiles were measured in each individual frame and also for a median stack of all frames in each night that showed the comet well separated from the negative imprint from the reference frame, and from any residuals due to saturated stars. The resulting profiles are shown in fig.~\ref{sbps}. This analysis also confirms that there was no detectable activity at 4.4 AU in September 2007. The October profile also appears star-like, while the November profile matches the image PSF in the inner parts. The presence of residuals from saturated stars near to the comet's position in November makes the profile difficult to measure at larger distances, so it is difficult to draw firm conclusions from it. By March 2008 the activity is obvious in the profile, as expected. 

Taken together, the photometry and profiles suggest that 67P's activity reached a detectable level from the ground at around 4.3 AU in November 2007, just as it was lost from view from Earth due to decreasing solar elongation. It is unfortunate that the two nights in November that were therefore critical presented the greatest challenges in data reduction, and that the night of the 15th produced no usable data to confirm the brightness increase seen on the night of the 12th. The presence of nearby stellar residuals in the profile indicates that we should be careful with the photometric identification of activity in the November data, but the seeing was good that night (FWHM $< 0.8\arcsec$) and the narrow aperture selected is well clear of any stars. 67P does not appear to have a sudden start in activity (a rapid rise in brightness) as seen in comet 10P/Tempel 2 \citep{Knight2012}, rather a more continuous increase in dust production that slowly crosses our detection threshold.

We therefore expect, assuming that the comet repeats its activity from one orbit to another, that detectable activity will begin in the current orbit at approximately 4.3 AU inbound, in March 2014. This is very early in the {\sl Rosetta} mission, during the phase when the spacecraft and instruments will be recommissioned following deep-space hibernation, and {\sl Rosetta} is still $\sim 5\times10^6$ km from the comet. This implies that an extra effort will be required by the {\sl Rosetta} team to get images of the comet as early as possible, in order to observe the start of activity. At this time the comet will be just returning to visibility from the Earth, and will again be seen against a crowded background, making observations from ground-based telescopes challenging. We believe that these extra efforts (by both ESA and ground-based observers) are justified as we expect activity to begin at large distance. It will also be important to test the sensitivity of ground-based observations against {\it in situ} measurements from {\sl Rosetta}'s instruments -- this mission will give us an interesting opportunity to test just how active a comet must be before careful processing of data from the largest telescopes can reveal the dust from Earth.

\subsection{The colour of the coma at large heliocentric distance}

Images were taken in 3 filters (Bessell $V$, $R$ and $I$) during March 2008, which we use to measure the colour of the comet. As 67P was weakly active there was significant flux from the coma. Using the method of \citet{Jewitt+Danielson84}, which assumes a steady state coma, we find that the coma made up $\sim20\%$ of the flux within an aperture of radius 10,000 km at the comet (equivalent to 3.7\arcsec{} at the time of observation). The colours of the comet within this aperture are found to be $(V-R) = 0.45 \pm 0.05$ and $(R-I) = 0.39 \pm 0.04$, which are redder than solar colours ($(V-R)_\sun = 0.35$; $(R-I)_\sun=0.33$) and similar to the average for cometary nuclei \citep{Snodgrass08,Lamy+Toth09}. The coma is slightly bluer than the nucleus, which has an average $(V-R) = 0.53 \pm 0.02$ and $(R-I) = 0.48 \pm 0.02$ \citep{Lamy06,Tubiana08,Tubiana11,Lowry12}. This implies that the grains in the coma are smaller than the dominant surface grains, preferentially scattering bluer wavelengths. An alternative explanation is the presence of significant gas emission lines, which can increase the flux in the $V$-band, but the $(R-I)$ colour is not strongly affected by gas emissions and is also seen to be bluer than the nucleus, implying that this is a grain size effect.

%%%%%%%%%%%%%%
%%% ARCHIVE BIT
%%%%%%%%%%%%%

\section{Archival data results}

\subsection{Heliocentric lightcurve}\label{hlc}

\begin{table*}
\caption{Photometry from archival data}

\small

\begin{center}
\begin{tabular}{l l c c c|c c|c c|c c l}
\hline
Date & Inst. & N\tablefootmark{a} & r\tablefootmark{b} & $\Delta T$\tablefootmark{c} & $m_R$ & $m_R$(r,1,0)\tablefootmark{d} & $m_R$ & $m_R$(r,1,0) &$Af\rho$\tablefootmark{e} & slope & calib.\tablefootmark{f} \\
 & & & (AU) & (days) &  \multicolumn{2}{c}{($\rho$ = 10,000 km)} & \multicolumn{2}{c}{($\rho$ = 5\arcsec)} &(cm) & & \\
\hline
\multicolumn{2}{l}{\it 1996 apparition:}\\
25/06/1995 & EFOSC & 3  & -2.50 & -205  & 17.46  & 15.61 & 17.60  & 15.75 & 46  & -1.93 & APASS \\
12/12/1995 & EMMI  & 1  & -1.37 & -36   & 13.15  & 12.22 & 13.87  & 12.94 & 317 & -1.28 & APASS \\
\hline
\multicolumn{2}{l}{\it 2002 apparition:}\\
09/09/2002 & KISO  & 13 &  1.32 & 22    & 13.34  & 11.45 & 13.94  & 12.05 & 598 & -1.90 & SDSS \\
11/12/2002 & SDSS  & 1  &  1.84 & 115   & 15.35  & 13.70 & 15.97  & 14.31 & 146 & -1.29 & SDSS \\
10/02/2003 & EFOSC & 1  &  2.30 & 177   & 15.57  & 14.57 & 16.39  & 15.39 & 103 & -1.21 & SDSS \\
08/03/2003 & EFOSC & 10 &  2.49 & 203   & 15.91  & 14.94 & 16.68  & 15.70 & 86  & -1.21 & SDSS \\
25/03/2003 & EMMI  & 2  &  2.61 & 220   & 16.61  & 15.30 & 17.26  & 15.95 & 67  & -1.21 & SDSS \\
30/04/2003 & FORS  & 5  &  2.86 & 256   & 18.08  & 15.97 & 18.38  & 16.27 & 44  & --	& SDSS \\
01/05/2003 & FORS  & 5  &  2.87 & 257   & 18.06  & 15.93 & 18.35  & 16.22 & 45  & -1.00 & SDSS \\
02/05/2003 & FORS  & 8  &  2.87 & 258   & 18.10  & 15.95 & 18.38  & 16.23 & 45  & -0.70 & SDSS \\
03/05/2003 & FORS  & 9  &  2.88 & 259   & 18.13  & 15.96 & 18.40  & 16.24 & 45  & -1.15 & SDSS \\
04/05/2003 & FORS  & 7  &  2.89 & 260   & 18.18  & 16.00 & 18.44  & 16.26 & 43  & --	& SDSS \\
03/06/2003 & FORS  & 2  &  3.08 & 290   & 18.93  & 16.23 & 18.83  & 16.13 & 40  & -0.82 & SDSS \\
04/06/2003 & FORS  & 1  &  3.09 & 291   & 19.20  & 16.49 & 19.11  & 16.40 & 32  & -1.24 & SDSS \\
19/06/2003 & FORS  & 2  &  3.19 & 306   & 19.42  & 16.51 & 19.18  & 16.27 & 33  & -0.94 & SDSS \\
20/06/2003 & FORS  & 1  &  3.19 & 307   & 19.42  & 16.50 & 19.18  & 16.26 & 33  & -1.07 & SDSS \\
23/06/2003 & FORS  & 1  &  3.21 & 310   & 19.50  & 16.55 & 19.23  & 16.27 & 32  & -1.59 & SDSS \\
24/06/2003 & FORS  & 1  &  3.22 & 311   & 19.47  & 16.50 & 19.19  & 16.23 & 34  & -1.41 & SDSS \\
23/02/2004 & SUSI  & 4  &  4.47 & 555   & 21.50  & 18.27 & 21.10  & 17.87 & 13  & -2.21 & ESO (n) \\
29/04/2004 & EFOSC & 5  &  4.73 & 621   & 21.65  & 18.76 & 21.56  & 18.67 & 9   & -2.38 & SDSS \\
16/06/2004 & FORS  & 10 &  4.89 & 669   & 22.42  & 19.05 & 22.04  & 18.67 & 7   & -2.60 & Tubiana \\
\hline
\multicolumn{2}{l}{\it 2009 apparition:}\\
30/05/2006 & FORS  & 61 & -5.61 & -1005 & 22.68  & 19.36 & 22.65  & 19.32 & 7   & -2.64 & ESO (n) \\
17/08/2006 & FORS  & 3  & -5.51 & -926  & 23.13  & 19.32 & --	  & --	  & 7   & --	& Tubiana \\
27/03/2008 & FORS  & 7  & -3.39 & -338  & 20.38  & 17.19 & 20.29  & 17.10 & 20  & --	& Landolt \\
29/03/2008 & FORS  & 7  & -3.38 & -336  & 20.47  & 17.30 & 20.57  & 17.40 & 18  & -2.55 & Landolt \\
30/03/2008 & FORS  & 7  & -3.38 & -335  & 20.30  & 17.13 & 20.28  & 17.11 & 21  & -2.50 & Landolt \\
31/05/2008 & FORS  & 8  & -2.98 & -273  & 18.90  & 16.54 & 18.93  & 16.57 & 28  & -1.97 & Landolt \\
03/06/2008 & FORS  & 8  & -2.96 & -270  & 18.76  & 16.45 & 18.80  & 16.50 & 30  & -1.91 & Landolt \\
04/06/2008 & FORS  & 8  & -2.95 & -269  & 18.78  & 16.49 & 18.83  & 16.54 & 29  & -1.89 & Landolt \\
04/07/2008 & FORS  & 6  & -2.75 & -239  & 17.77  & 16.10 & 17.90  & 16.23 & 36  & -1.92 & Landolt \\
10/08/2008 & FORS  & 5  & -2.48 & -202  & 16.16  & 15.22 & 16.51  & 15.56 & 66  & -1.72 & Landolt \\
03/09/2008 & FORS  & 16 & -2.30 & -178  & 16.23  & 15.22 & 16.58  & 15.57 & 56  & -1.70 & ESO (n) \\
05/09/2008 & FORS  & 16 & -2.28 & -176  & 15.90  & 14.87 & 16.27  & 15.24 & 77  & -1.79 & Landolt \\
06/09/2008 & FORS  & 16 & -2.28 & -175  & 15.96  & 14.93 & 16.32  & 15.28 & 73  & -1.62 & Landolt \\
07/09/2008 & FORS  & 16 & -2.27 & -174  & 16.21  & 15.17 & 16.56  & 15.51 & 58  & -1.63 & ESO (n) \\
13/09/2008 & FORS  & 1  & -2.22 & -168  & 15.94  & 14.84 & 16.27  & 15.17 & 75  & -1.77 & Landolt \\
21/10/2008 & FORS  & 5  & -1.93 & -130  & 15.99  & 14.47 & 16.31  & 14.80 & 79  & -1.71 & ESO (n) \\
25/10/2008 & FORS  & 5  & -1.90 & -126  & 15.81  & 14.26 & 16.16  & 14.60 & 93  & -1.85 & ESO (n) \\
26/10/2008 & FORS  & 5  & -1.89 & -125  & 16.01  & 14.45 & 16.33  & 14.76 & 78  & -1.86 & ESO (n) \\
25/01/2009 & LOT   & -- & -1.31 & -34	& 13.52  & 11.69 & --	  & --	  & 476 & --	& Lara \\ 
28/01/2009 & EFOSC & 3  & -1.30 & -31   & 13.62  & 11.79 & 13.97  & 12.13 & 426 & -1.65 & ESO (d) \\
30/01/2009 & LOT   & -- & -1.29 & -29	& 13.46  & 11.62 & --	  & --	  & 490 & --	& Lara \\ 
31/01/2009 & LOT   & -- & -1.29 & -28	& 13.46  & 11.63 & --	  & --	  & 486 & --	& Lara \\ 
23/02/2009 & LOT   & -- & -1.25 & -5	& 13.05  & 11.21 & --	  & --	  & 666 & --	& Lara \\ 
25/02/2009 & LOT   & -- & -1.25 & -3	& 13.37  & 11.53 & --	  & --	  & 497 & --	& Lara \\ 
26/02/2009 & LOT   & -- & -1.25 & -2	& 12.97  & 11.12 & --	  & --	  & 721 & --	& Lara \\ 
12/03/2009 & LOT   & -- & 1.26  & 12	& 12.87  & 11.01 & --	  & --	  & 813 & --	& Lara \\ 
19/03/2009 & CAFOS & -- & 1.27  & 19	& 13.04  & 11.16 & --	  & --	  & 724 & --	& Lara \\ 
15/02/2010 & EFOSC & 5  &  3.49 & 352   & 19.93  & 17.34 & 19.86  & 17.27 & 19  & -1.88 & SDSS \\
16/02/2010 & EFOSC & 3  &  3.49 & 353   & 20.02  & 17.44 & 19.96  & 17.38 & 17  & -2.29 & SDSS \\
14/07/2010 & EFOSC & 3  &  4.26 & 501   & 21.74  & 18.30 & 21.15  & 17.72 & 11  & --	& SDSS \\
\hline
\end{tabular}
\end{center}
\tablefoottext{a}{Number of images used.}
\tablefoottext{b}{Heliocentric distance. Negative numbers indicate pre-perihelion, positive post-perihelion.}
\tablefoottext{c}{Time before or after perihelion date.}
\tablefoottext{d}{Magnitude reduced to unit geocentric distance and zero phase angle.}
\tablefoottext{e}{Using $\rho$ = 10,000 km at the comet, and corrected to zero phase angle.}
\tablefoottext{f}{Calibration source for each night: APASS / SDSS - field stars from these catalogues; ESO - nightly (n) or default (d) zeropoints from ESO web page; Landolt - zeropoint calculated using observations of Landolt/Stetson standard stars on the same night; Tubiana - zeropoint solutions taken from \citet{Tubiana-thesis} (from Landolt stars); Lara - photometry calculated from $Af\rho$ values reported in \citet{Lara11}. The $Af\rho$ value quoted here has been corrected to zero phase (Lara measurements were all taken at $\sim36\degr{}$). }
\label{archive-results}
\end{table*}%

\begin{figure}
   \centering
   \includegraphics[width=\columnwidth]{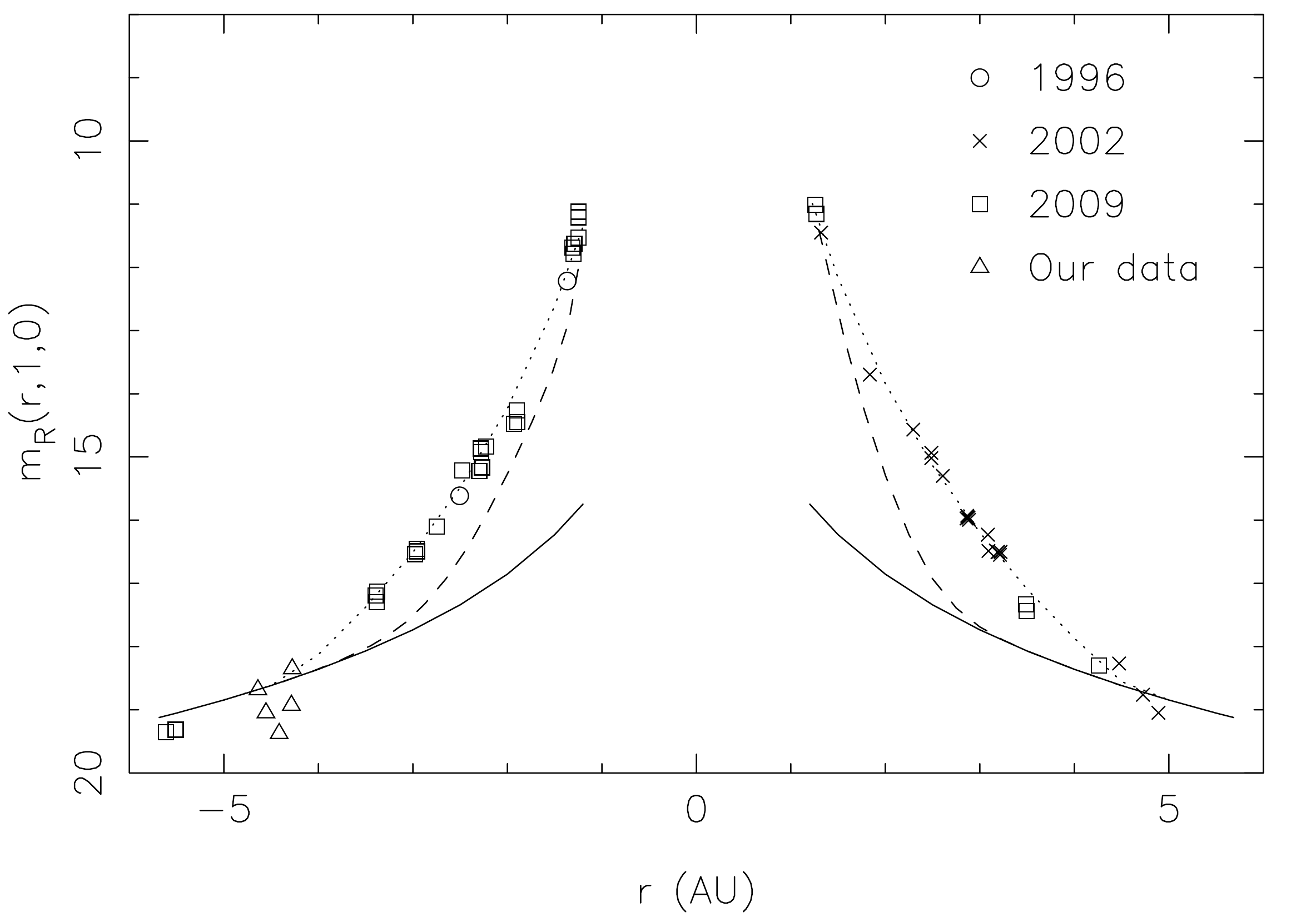} 
   \caption{Heliocentric lightcurve, showing the $R$-band magnitude measured within an aperture of radius 10,000 km at the distance of the comet, corrected to unit geocentric distance and zero phase angle, against heliocentric distance in AU. Negative heliocentric distance shows pre-perihelion measurements. Different symbols show data taken during different apparitions, and highlight `our data' presented in section \ref{activity_start} (which were taken at the start of the 2009 apparition). The solid black line shows the expected magnitude of the bare nucleus, based on the absolute magnitude reported by \citet{Tubiana11}. The dashed line shows a prediction for the total magnitude based on a water production rate model (see section \ref{gas-section}), while the dotted lines show a power law fit to the data.}
   \label{fig:heliocentric}
\end{figure}

\begin{figure}
   \centering
   \includegraphics[width=\columnwidth]{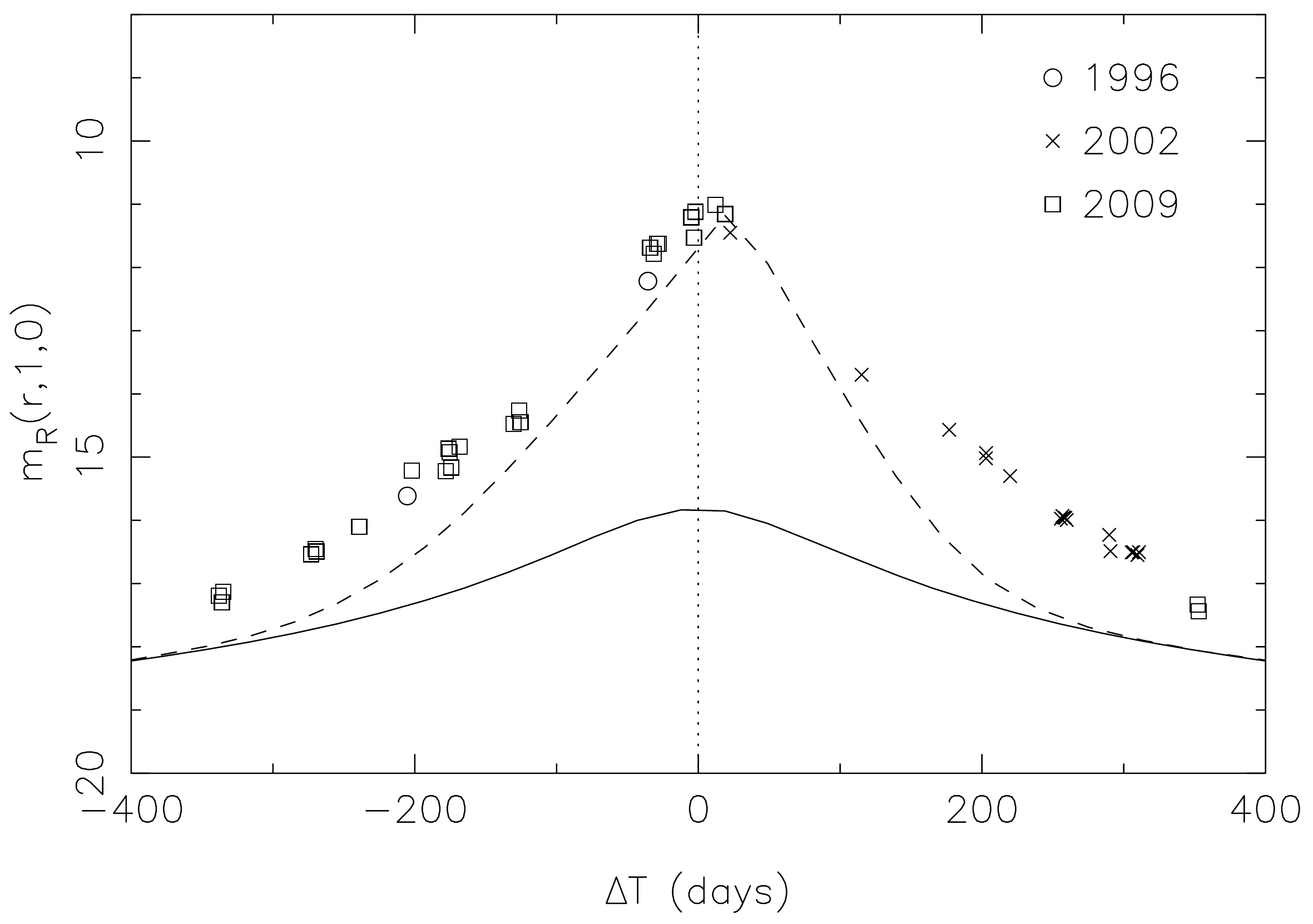} 
   \caption{Same as fig.~\ref{fig:heliocentric}, but with heliocentric lightcurve against time from perihelion in days. In this plot we show only data within $\pm$ 1 year of perihelion, to highlight the peak region.}
   \label{fig:time}
\end{figure}

Heliocentric lightcurves give a good overview of the activity level of a comet around its orbit. The series of papers by Ferr{\'i}n \citep[e.g.][]{Ferrin,Ferrin-atlas} have presented heliocentric (or `secular') lightcurves for a number of comets based on photometry reported to the Minor Planets Center or available in the literature, and give a general description of the behaviour of each comet, even if the photometric scatter in these plots is large due to the varied data sources. 

Here we present a heliocentric lightcurve for 67P based on a consistent re-reduction of professional data taken from archival sources. All data shown in figs.~\ref{fig:heliocentric} and \ref{fig:time}, and given in table~\ref{archive-results}, come from our own reprocessing of archival images, with the exception of some points around perihelion during the 2009 apparition taken from \citet{Lara11}, where the reported $Af\rho$ within $\rho=10,000$ km was converted back into a flux within this aperture.

We choose to display the lightcurve measured within a $\rho = 10,000$ km aperture for all measurements, to maintain a consistent physical volume of coma, independent of viewing geometry. We measured the photometry within a selection of different apertures, including fixed physical sizes and fixed apparent radii (we give results from both $\rho = 10,000$ km and $\rho=5\arcsec$ in table~\ref{archive-results}). The table also gives the $Af\rho$ quantity and slope of the radial profile of the comet (see section \ref{dust_production}). Using a fixed aperture means that no attempt is made to measure the `total' flux from the comet; near to perihelion the coma is clearly larger than 10,000 km in radius. This is quite deliberate, as a total flux measurement either requires a very large aperture to be sure to contain all possible coma (which would possibly mean an aperture larger than the field of view of the camera in some cases), and removal of all stars within that aperture, or to somehow define the edge of the coma, which clearly depends on the sensitivity of the camera used.

The measured photometry was reduced to geocentric distance $\Delta = 1$ AU and phase angle $\alpha=0\degr{}$ using 
\begin{equation}
m_R(r,1,0) = m_R -5 {\rm log} \Delta  - \beta\alpha
\end{equation} 
with the phase coefficient $\beta=0.02$ mag deg$^{-1}$, as found for cometary dust over a range of phase angles up to $\alpha \approx 30\degr{}$ \citep{Meech87}. An alternative phase function, which is non-linear at larger angles, combines  measurement of comet 1P/Halley \citep{Schleicher98} with calculations by \citet{Marcus07}\footnote{See {\tt http://asteroid.lowell.edu/comet/dustphase.html} for details.}. For $\alpha < 55\degr{}$ this is given by
\begin{equation}
m_R(r,1,0) = m_R -5 {\rm log} \Delta  - 2.5(0.01807\alpha - 0.000177\alpha^2),
\end{equation} 
which can approximated with a linear $\beta=0.04$ mag deg$^{-1}$ relationship over most phase angles observable from Earth. Using this instead of $\beta=0.02$ mag deg$^{-1}$ makes only a small difference to the resulting photometry (increasing the perihelion brightness slightly, but not noticeably changing the shape of the lightcurve). We choose to use the simple linear phase function, and return to this topic in the discussion (see section \ref{phase-funct}).

The phase function behaviour of cometary dust is quite different from that of nuclei. The nucleus phase darkening has been measured for only a handful of comets, but these can be described by linear functions with an average coefficient of $\beta = 0.053 \pm 0.016$ mag deg$^{-1}$ \citep{Snodgrass11}. Slopes of $\beta = 0.059 - 0.076$ mag deg$^{-1}$ have been found for 67P \citep{Lowry12,Tubiana11}. These are considerably steeper than the dust function, so the choice of `dust-like' or `nucleus-like' phase correction can have a significant effect on the photometric results. As the heliocentric lightcurve primarily contains data from the active phases, where the flux from the coma dominates, we use the dust law for all data. The difference for the handful of observations of the inactive nucleus is small, as these were generally taken near to opposition. $Af\rho$ values are also corrected to zero phase.

The lightcurve demonstrates a number of noteworthy features. Firstly, the lightcurve is remarkably `clean', showing the advantage of using only data taken with large professional telescopes and consistent methods. We conservatively estimate the uncertainty on any given point to be $\sim 0.1$ mag.~(the formal uncertainties including photon noise and calibration are typically $\sim 0.01$ mag., but do not include any systematic uncertainty due to either the relatively `quick and dirty' flat-fielding or occasional faint stars within the photometric aperture). 
This gives us some confidence in making further predictions based on the lightcurve. Power law fits to the data describe the flux as $\propto r^{-5.2}$ pre-perihelion and $\propto r^{-5.4}$ post-perihelion.
The second point to note is that data taken around different perihelion passages (1996, 2002 and 2009) all fit on the same smooth curve, implying that the activity level of the comet does not vary significantly between apparitions, and that any predictions for 2014/5 may be of some use (of course, with the usual caveats that even `well behaved' comets can have unpredictable moments). It is unfortunate that the section of the lightcurve between -4 and -3.5 AU is one of the few badly covered parts, so we cannot add further to the question of the turn-on time discussed in the previous section, beyond noting that the pre-perihelion slope is consistent with departure from a bare nucleus at 4.3 AU. There was only one additional observation in this period in the archive, from the University of Hawaii 2.2m in August 2001 ($r = -3.5$ AU), where the comet was seen but was merged with brighter background stars. As this data set was a $BVRI$ snap-shot, with only one frame per filter, it is not possible to apply DIA techniques to recover the $R$-band photometry. Finally, we note that the overall shape of the lightcurve is reasonably symmetrical around perihelion, although the comet is consistently brighter (by around 1 magnitude) at the same distance post-perihelion when compared to pre-perihelion. This is more clearly seen when the heliocentric lightcurve is plotted against time from perihelion (fig.~\ref{fig:time}), which shows that the brightness peaks in the weeks immediately after perihelion.

The idea that 67P reaches its peak in activity shortly after perihelion is in agreement with past results based on water production rate \citep{Schleicher} and coma morphology \citep{Vincent13}. Lightcurves based on amateur photometry around perihelion also show this asymmetry \citep{Kidger03,Ferrin}, with the peak occurring approximately one month after perihelion. The asymmetry can be explained by different hypotheses. Firstly, it can be due to the time taken for the thermal wave to travel from the comet's surface to the buried ice, meaning that the maximum sublimation rate lags behind the maximum solar insolation. Alternatively, the production of dust and gas could be a two stage process: Material is mostly lifted in large grains which subsequently fragment, delaying the majority of the production rate. This scenario is unlikely to explain a lag of weeks though, as the timescale for material to pass out of the 10,000 km radius aperture is only 2 or 3 days at typical grain speeds. Finally, the activity peak could be due to a seasonal effect. Based on the pole orientation measured by \citet{Vincent13} and \citet{Lowry12}, it is expected that the comet reaches equinox around 50 days before perihelion, so it is possible that there are areas near the previously unlit pole (e.g. in crater-like depressions) that only see sunlight near to perihelion (and hence produce a boost in activity at this time). \citet{Vincent13} use such a seasonal model to explain the relative strengths of jets seen in the coma. Observations by {\sl Rosetta} will allow us to differentiate between these various effects.

\subsection{Dust production}\label{dust_production}

The $R$-band observations used in measuring the heliocentric lightcurve are sensitive to reflected sunlight from dust in the coma, and are not significantly affected by emission lines from gas. It is therefore natural to use these measurements to constrain the amount of dust in the coma, and the rate at which the comet produces dust. The dust brightness for active comets is often quantified by the $Af\rho$ parameter \citep{Ahearn89}. The conversion from surface brightness to mass loss rate requires assumptions on the albedo, density, velocity and size distribution of dust grains, all of which are only poorly constrained. Various authors have built complex dust models which fit these parameters to the observed coma brightness and morphology \citep[e.g.][]{Agarwal07,Agarwal10,Fulle10}. The recent paper by \citet{Fink+Rubin12} calculates expected $Af\rho$ values considering variable dust albedo and phase behaviour dependent on grain size, based on the dust flux simulations for 67P by \citet{Tenishev2011}.
This sort of modelling is beyond the scope of this paper; instead we report the observed $Af\rho$ and briefly consider some caveats on its use.

$Af\rho$ is useful as it is easily calculated and widely employed, providing a first order comparison of activity levels between different comets and at varying heliocentric distances. The values we derive for 67P are typical for Jupiter family comets, of order $10^2 - 10^3$ cm around perihelion. The use of $Af\rho$ to compare between comets (or even other observations of 67P) does have to be treated with care though. Firstly, the values we give in table~\ref{archive-results} use a $\beta = 0.02$ mag deg$^{-1}$ phase function to correct them to the values that would have been observed at a constant phase angle. We correct to $\alpha=0\degr$, while others normalise to other angles or do no not apply any phase function correction. Secondly, it is important to note that $Af\rho$ is independent of wavelength and of the choice of aperture radius $\rho$ only for an idealised coma with grey dust and a $1/\rho$ brightness profile. In real comets this is not always the case, as the particle size distribution can influence the coma colour, and the assumption of a $1/\rho$ profile requires a steady state (dust being produced from the nucleus at the same rate it flows out of the aperture). We minimise these uncertainties by using a fixed physical aperture size and observations at a fixed wavelength, and a consistent correction for phase, so the $Af\rho$ values we find can be compared with each other to assess dust production around the orbit. To judge how well our $Af\rho$ values can be used in comparison with other work, we test how well 67P meets the steady state assumption.

\begin{figure*}
   \centering
   \includegraphics[width=\textwidth]{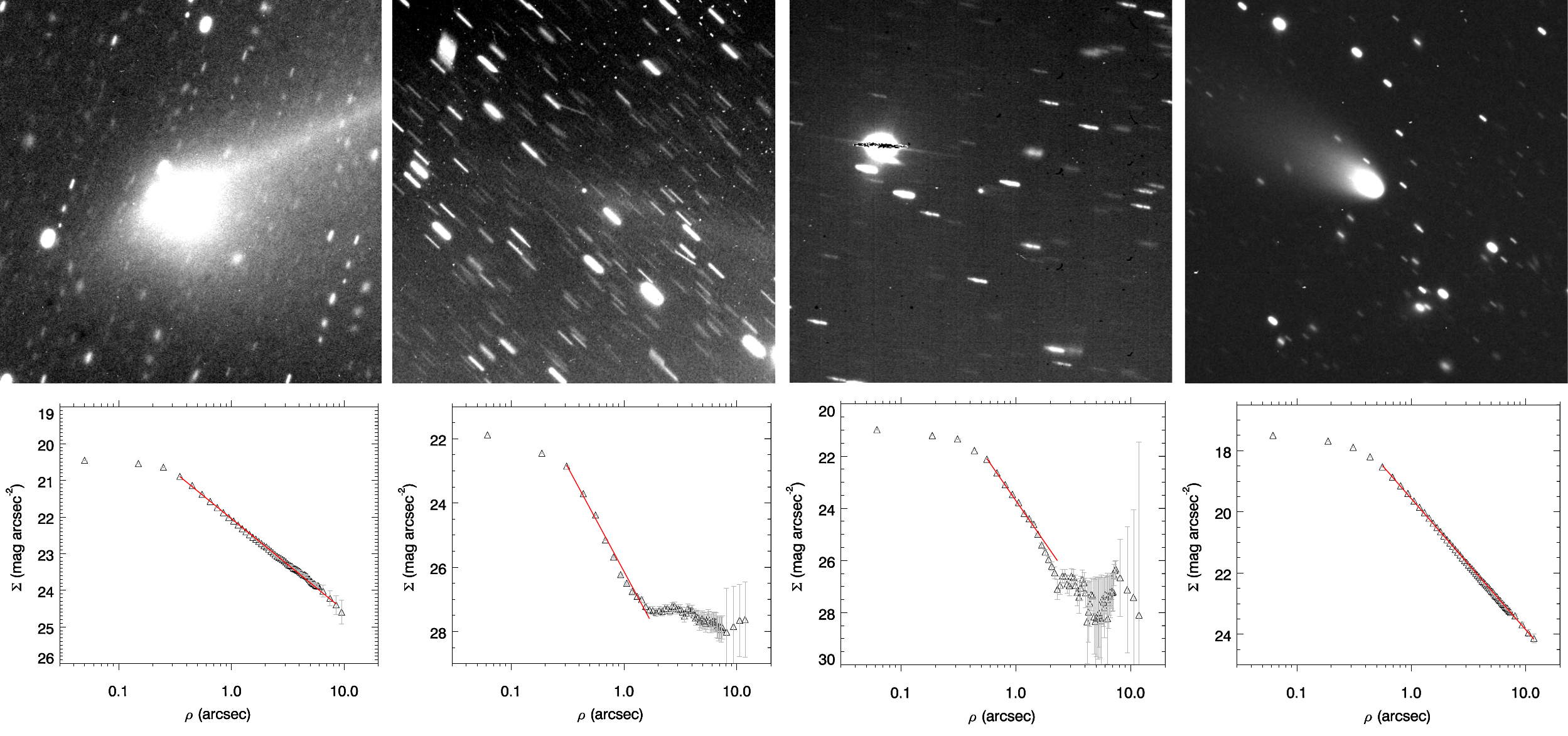} 
   \caption{Representative images and corresponding surface brightness profiles for 67P at four different points in its orbit. From left to right: May 2003, at 2.9 AU post-perihelion, showing high activity and a trail or neck-line of larger particles; June 2004, 4.9 AU post-perihelion, with the nucleus inactive but a clear neck-line; March 2008, 3.4 AU, at the start of visible activity; October 2008, 1.9 AU pre-perihelion, displaying a broad tail.}
   \label{fig:profiles}
\end{figure*}

It is clear from images of the comet (representative examples are shown in fig.~\ref{fig:profiles}) that the morphology of its coma and tails changes throughout its orbit. Pre-perihelion the coma is quite symmetric, with a broad fan developing into a dust trail, while post-perihelion larger particles form a `neckline'. Further image processing reveals jets within the coma, which are seen to change in strength around the orbit. \citet{Agarwal07} and \citet{Vincent13} discuss the large scale and fine structure morphology of the comet in detail. 

\citet{Tozzi} found that $Af\rho$ varied with $\rho$ for observations of 67P at $r=2.3$ AU pre-perihelion, and so the coma did not behave as $1/\rho$. We find that 67P does not match the steady state assumption over a wider range in heliocentric distance, by measuring the slope of the radial profile in all the archival data sets. When plotted as magnitudes (or fluxes on a logarithmic scale) against log($\rho$), the surface brightness profile of an active comet should have a linear profile with a slope of -1, or -1.5 if the effects of radiation pressure are taken into account \citep{Jewitt+Meech87}. We find that the slopes vary between shallower than -1 to -2 or steeper, with a correlation between slope and heliocentric distance. The steepest slopes are meaningless, as they correspond to images when the comet is inactive (fig.~\ref{fig:profiles}), and the profile shape is actually the image PSF (and therefore a linear fit to a gaussian profile, with the result dependent on the seeing), but there is still a correlation when only the active profiles are considered. A possible explanation for this effect is the longer time that larger dust grains remain near to the nucleus, acting to slow the overall rate at which dust leaves an aperture when there are more large grains (presumably when activity is highest near to perihelion). A more complex dust model needs to be employed to look at this effect in detail, so it is not considered further here.

The comet's changing morphology and variable surface brightness profile show that $Af\rho$ measurements must be treated with caution. Despite this caveat, we find that inspection of the values determined from the archival data can give some useful constraints.
We measure $Af\rho$ values with a peak at $\sim 1000$ cm in the weeks after perihelion, with values $\sim 100$ cm at $r = 2$ AU, $\sim 50$ cm at 3 AU, and dropping to single figures (where the flux is entirely from the inactive nucleus, and $Af\rho$ is consequently meaningless) beyond 4.5 AU. $Af\rho$ is larger at a given distance post-perihelion than it is pre-perihelion, consistent with the slight offset in the brightness peak and also with the observed tendency of comets to remain active to larger distances outbound \citep[e.g.][]{Kelley-SEPPCON}. All measurements are consistent with single power-law fits pre- and post-perihelion, given by
\begin{equation}\label{afro-r-eqn}
Af\rho = \ 958 \times r^{-3.18} \quad \textrm{(cm)}
\end{equation}
and
\begin{equation}
Af\rho  = 1552 \times r^{-3.35} \quad \textrm{(cm)}
\end{equation}
respectively. These are steeper than the canonical $1/r^2$ often assumed for comets (and the average $Af\rho \propto r^{-2.3}$ found by \citet{Ahearn95}), but are shallower than the $r^{-5.8}$ fit for 67P found by \citet{Kidger03}. \citet{Schleicher} found exponents of $-9.5 \pm 5.6$ and $-8.0 \pm 3.6$ pre-perihelion, and $-0.6 \pm 1.3$ and $-0.4 \pm 1.2$ post-perihelion, for $Af\rho$ in UV and green continuum filters, but over a much reduced range in $r$ (especially pre-perihelion). \citet{Ahearn95} found a slope of $-1.34 \pm 0.81$, for the same range of distances ($1.3 < r < 1.9$ AU) post-perihelion. Our result is significantly shallower than the best fit found by \citet{Agarwal07}, who obtained $Af\rho = 2393 \times r^{-5.08}$ cm by fitting phase corrected values from a variety of sources (including both the \citet{Kidger03} and \citet{Schleicher} data sets) in the range $-2 < r < 3$ AU (over three apparitions), although these authors note that the data have a large scatter and the obtained exponent is unexpectedly steep.

$Af\rho$ is, under the simplest assumptions, directly proportional to the dust production rate, with an empirical calibration suggesting that $Af\rho$ in cm $\approx$ $Q_{\rm d}$ in kg s$^{-1}$ \citep{Ahearn95}. We can therefore compare the $r$ dependence we find to that assumed ($Q_{\rm d} \propto r^\delta$) in various dust models: 
\begin{enumerate}
\item \citet{Agarwal10} assume an equivalent of $\delta = -5$, based on the \citet{Agarwal07} fit to $Af\rho$. 
\item \citet{Kelley08,Kelley09} use $\delta = -5.8$, from the fit by \citet{Kidger03}. 
\item \citet{Ishiguro08} uses the neck-line appearance to find a weak constraint on $\delta = -3$, although this model considers only activity near perihelion. 
\item The model by \citet{deSanctis05} can be approximated pre-perihelion by $\delta=-3.5$ for $r<2$ AU, with a much steeper $\delta = -10.2$ beyond this (based on their fig.~9, model A). Post-perihelion we find that $\delta=-3.2$ for $r<2$ AU and $\delta = -10.5$ for $r>2$ AU fit their model.
\item The complex GIADA dust model \citep{Fulle10} does not provide a simple scaling with $r$ for comparison, as it is based on a higher order fit to $Af\rho$ data. It is not clear that such a function is required, given the error bars on \citet{Fulle10}'s fig.~1; we find that power laws with $-3 < \delta < -2$  given reasonable fits to the displayed data, although these do not necessarily trace the complex behaviour of the model. 
\end{enumerate}
Most of these models agree on a perihelion dust production rate $Q_{\rm d} \approx 1000$ kg s$^{-1}$ (with the exception of \citet{Ishiguro08}, who obtain an order of magnitude lower rate), comparable to the number from a simple calibration of $Af\rho$. It is clear that the generally steeper dependence on $r$ in the models means that they predict relatively little dust at larger distances than our photometry suggests.  While it would certainly be worth applying more complex models to the data we present here, it appears that current models underestimate the dust production that can be expected at large heliocentric distance, and {\sl Rosetta} should be prepared to encounter significant dust early in its mission.

\subsection{Gas production}\label{gas-section}

For {\sl Rosetta}, it is important to constrain the gas production rate of 67P around the orbit. This is actually of greater importance than the dust production rate, in terms of mission planning, as the effects of gas drag on the large solar panels present more problems (and transfer more momentum) than dust impacts on the spacecraft. 
The gas production rate is, however, more difficult to measure, requiring spectra or images in narrow-band filters, and is typically only possible when the comet is near to the Earth and Sun, and therefore already highly active. The $R$-band images we present are not sensitive to gas emissions, but can be used to compare predictions on total brightness from gas production models.

The most substantial set of gas production rate observations come from \citet{Schleicher}, who presents narrow-band photometry taken around the 1982 and 1996 perihelion passages, reaching a maximum distance of $r=1.86$ AU outbound, and also reviews measurements from other authors. 
\citet{Ootsubo12} measured water and CO$_2$ production rates using the {\sl Akari} satellite when 67P was at $r=1.8$ AU inbound in 2008.
There is considerable scatter in these measurements, attributed to rotational variability by \citet{Schleicher}, but in general water production rate is found to peak at $Q \approx 10^{28}$ molecules s$^{-1}$ around a month after perihelion. 

For {\sl Rosetta} planning there are agreed minimum and maximum activity levels expected, in terms of production rates for H$_2$O, CO and CO$_2$, at perihelion, 2, 3 and 3.5 AU \citep{Biele-ESA-note,Biele+Ulamec13}. These define boundary conditions for planning trajectories (and science operations) possible in the so called `Low' and `High' activity cases, and are based on simple extrapolations from the observations rather than any complex model.
The {\sl ALICE} and {\sl OSIRIS} instruments on {\sl Rosetta} are using a power law fit to the observed water production rates for their planning, based on time from perihelion $\Delta T$ rather than $r$ to include the asymmetric peak in activity (A'Hearn, private communication). It is given by 
\begin{eqnarray}\label{Mike-fit}
\textrm{log}(Q) = 27.66 + 0.006836\Delta T, & \Delta T < 35\\
\textrm{log}(Q) = 28.10 - 0.009858\Delta T, & \Delta T > 35\notag
\end{eqnarray}
This fit falls neatly between the minimum and maximum activity levels considered by ESA, and so, while obviously only an approximate model, provides a reference production rate for the medium activity case, which we believe to be the most likely actual scenario at the comet. 

The fit is displayed for comparison with our photometry in fig.~\ref{fig:heliocentric} and fig.~\ref{fig:time}, where we have converted water production to total magnitude using the empirical relationship log($Q$) = 30.675 - 0.2453$m_h$ \citep{Jorda92,Jorda08}. The total magnitude $m_h$ is for visual observations (assumed to approximately match the $V$-band), and is reduced to unit geocentric distance but does not include any phase function correction. As it is based on estimates of `total' brightness, it is not based on observations through any fixed aperture, with individual measurements showing great variation. The relationship is essentially uncalibrated for distances beyond 2 - 3 AU, at the limit of the measurements on $Q$(H$_2$O) for normal comets. Together, these limitations mean that the magnitudes estimated from this method are indications only, and not more accurate than $\pm 0.5$ magnitudes, but they do still allow a useful comparison. Fig.~\ref{fig:heliocentric} shows that the total brightness estimated from the water production rate is significantly lower than the measured $R$-band magnitude (including a correction of $(V-R)=0.5$), despite the fact that our measurements are taken within $\rho = 10,000$ km, and therefore underestimate the total brightness near to perihelion. The difference is  more than a magnitude at all distances pre-perihelion, only giving similar results around the peak in water production. As both the magnitude and the plotted water model are based on observations, including data from the same perihelion passage, this is a real effect, not due to a difference in models. 

This implies that 67P must be a relatively dusty comet, either due to water lifting more dust from the surface than other comets with the same production rate, or due to a significant component of the activity being driven by a different gas species (e.g. CO or CO$_2$). The measurement by \citet{Ootsubo12} shows the ratio of CO$_2$/H$_2$O for 67P to be relatively low (7\%) compared with other comets (median 17\%) measured within 2.5 AU from the Sun. According to the taxonomy of \citet{Ahearn95} 67P is a carbon-chain depleted comet (although only `mildly depleted' based on the reanalysis of the same data by \citet{Schleicher}), meaning that it has a lower C$_2$/CN ratio than other comets. As the parent species of C$_2$ and CN are still debated, this does not tell us much about what is driving the dust. The production rates of other gasses (C$_3$, NH) relative to OH are close to the average values for the depleted group of comets \citep{Ahearn95}. Surprisingly, \citet{Ahearn95} also find a typical `dust production' for 67P, with log($Af\rho$/$Q$(OH)) = $-25.23\pm0.50$ cm s molecule$^{-1}$, compared with the mean for depleted comets of $-25.30\pm0.29$. \citet{Schleicher} slightly update the value to -25.27, and note that this value is dustier than `typical' comets, and almost identical to that found for 81P/Wild~2. 
We find a mean value of log($Af\rho$/$Q$(OH)) = $-24.89\pm0.21$ cm s molecule$^{-1}$, based on the OH production rates from \citet{Schleicher} and using the $r$ dependency given above to generate the expected values of $Af\rho$ at the time of each observation. The difference is largely due to the fact that we have corrected our $Af\rho$ values to zero phase angle. Following the same approach we find an average dust/water relationship with log($Af\rho$/$Q$(H$_2$O)) = $-24.94\pm0.22$ cm s molecule$^{-1}$, where we have included all available measurements of the H$_2$O production rate. These measurements cover a range $-1.8 < r < 1.9$ AU, and there is a trend of increasing dust-to-gas ratio over this range (which was also noted by \citet{Schleicher}), so the average value must be treated with caution. The trend can be approximated by $Af\rho/Q \propto r^{2.8}$, implying that the dust-to-gas ratio increases with increasing distance from the Sun. Taken together with equation~\ref{afro-r-eqn}, this suggests that the pre-perihelion water production rate can be estimated by  
\begin{equation}\label{Q-r-eqn}
Q(\textrm{H}_2\textrm{O}) \approx 2.3\times10^{28} r^{-5.9},
\end{equation}
implying that {\sl Rosetta} will encounter production rates of $Q \sim 6\times10^{24}$ molecules s$^{-1}$ at $r = 4$ AU, $\sim 3\times10^{25}$ at 3 AU, and $\sim 4\times10^{26}$ at 2 AU. These must be treated as approximate, as it is unlikely that the $Af\rho \leftrightarrow Q$(H$_2$O) relationship is valid at these larger distances. The exponent is similar to that found by \citet{Schleicher} at smaller $r$, who obtained $Q$(OH) $\propto r^{-6.4}$ and $\propto r^{-5.4}$ for pre- and post-perihelion, respectively.

\subsection{Dust phase function}\label{phase-funct}

Finally, we attempt to put independent constraints on the phase function of the dust in 67P's coma. To do this we \emph{assume} that the lightcurve follows simple power laws, based on the good fit to the heliocentric lightcurve based on $\beta = 0.02$ mag deg$^{-1}$, and repeat the fitting process allowing $\beta$ to vary.
We continue to fit the pre- and post-perihelion power laws separately, with one power law for each, using a single value of $\beta$ for each pair of fits (i.e. we allow the heliocentric brightness dependence to be different on inbound and outbound legs as before, and assume that the dust phase function does not change around the orbit). By minimising the root-mean-squared variation around the best fit straight lines $m_R$ vs log($r$), we find an optimal $\beta = 0.03 \pm 0.02$ mag deg$^{-1}$. The $r$-dependence does not change much from that found assuming either $\beta = 0.02$ mag deg$^{-1}$ or the \citet{Schleicher98} phase function, which is not surprising as they bracket the best fit value, and both are covered by the formal uncertainty. Using $\beta = 0.03$ mag deg$^{-1}$ we find the $R$-band flux is $\propto r^{-5.4}$ pre-perihelion and $\propto r^{-5.6}$ post-perihelion.

\section{Activity model}\label{karen-models}

The simple power-law relationships given in previous sections are useful as they present a straightforward way to present \emph{observed} trends, and make predictions based on those. However, there is no physical reason to expect that the real outgassing rate actually follows such rules.
Ultimately,
full 3D thermal evolution models \citep[e.g.][]{Prialnik04} will be key to understanding
and interpreting the composition and activity of comets, but this is beyond the scope of this work.
As a first step we employ a simple ice sublimation model to explore
activity over a range of distances.  We use the model developed by \citet{Meech86} and applied to 103P/Hartley 2, based on the long campaign of photometry on that comet before and after the {\sl EPOXI} mission encounter \citep{Meech11}. The physics behind the model are described in more detail by \citet{Meech+Svoren04}, but are briefly introduced here.  

As ices sublimate, from either the surface or subsurface
layers, the escaping gas drags dust from the nucleus to escape into
the coma and tail.  In the absence of gas fluorescence, the total
comet brightness has a contribution from the nucleus and the scattered
light from the dust.  
It can be shown that the total coma brightness can be
expressed as a function of mass loss \citep{Meech86} via:
\begin{equation}
m_{coma} = 30.7 -2.5{\rm log}_{10}\left[\frac{p_{\lambda}(dM/dt) t}{D_{\rm g} a r^2 \Delta^2}\right]
\end{equation}
where the time, $t$, is a function of the projected
aperture size and grain velocity (for simplicity we assume the
Bobrovnikoff approximation $v \approx r^{-0.5}$ m s$^{-1}$, for $r$ in AU), and
$a$, $p_\lambda$ and $D_{\rm g}$ are the grain radius, albedo, and density, respectively.  The
mass loss is computed using the energy balance at the nucleus:
\begin{equation}
F_\odot (1-A)/r^2 = \chi [\epsilon \sigma T^4 + L(T)(dm_s/dt) + \kappa (dT/dz)].
\end{equation}
The left hand side of the equation is the incident solar flux and
the terms on the right hand side represent the blackbody energy,
the energy going into sublimation and conduction into the interior
(which we assume is negligibly small).  $\chi$ is a rotation
parameter expressing whether the heat is deposited only on one face
of the nucleus (slow rotator) or evenly over the whole surface (fast
rotator).  The mass loss per unit area, ($dm_s/dt$) is related to
the sublimation vapor pressure and the average speed of the
gas molecules leaving the surface.  The sources for the latent
heats $L(T)$ and sublimation vapor pressures for some common
ices are summarized in \citet{Meech86}.

The free parameters in the model include: ice type, nucleus radius,
albedo, emissivity, density, properties of the dust (sizes,
density, phase function), and fractional active area. For 67P we have the advantage that many of the required properties are known (or are at least reasonably well constrained), 
leaving us to adjust fractional
active area and ice composition to match the light curve. The assumed values for the model parameters are given in table \ref{Kmodel-table}.

\begin{table}
\caption{Parameters used in the activity model}
\begin{center}
\begin{tabular}{l l l}
\hline
Parameter & Value & Reference\\
\hline
{\it Nucleus:}\\
Phase coefficient & 0.076	mag deg$^{-1}$ & \citet{Tubiana11}\\
Albedo & 0.054 & \citet{Kelley09}\\
Radius & 2.04 km & \citet{Kelley09}\\
Emissivity & 0.9 & assumed\\
Rotation period	 & 12.68 hr (slow) & \citet{Tubiana11}\\
Density & 400 kg m$^{-3}$ & assumed\\
{\it Dust:}\\
Phase coefficient & 0.02 mag deg$^{-1}$ & assumed\\
Grain density & 1000 kg m$^{-3}$ & assumed\\
\hline
\end{tabular}
\end{center}
\label{Kmodel-table}
\end{table}%

\begin{figure}
   \centering
   \includegraphics[width=\columnwidth]{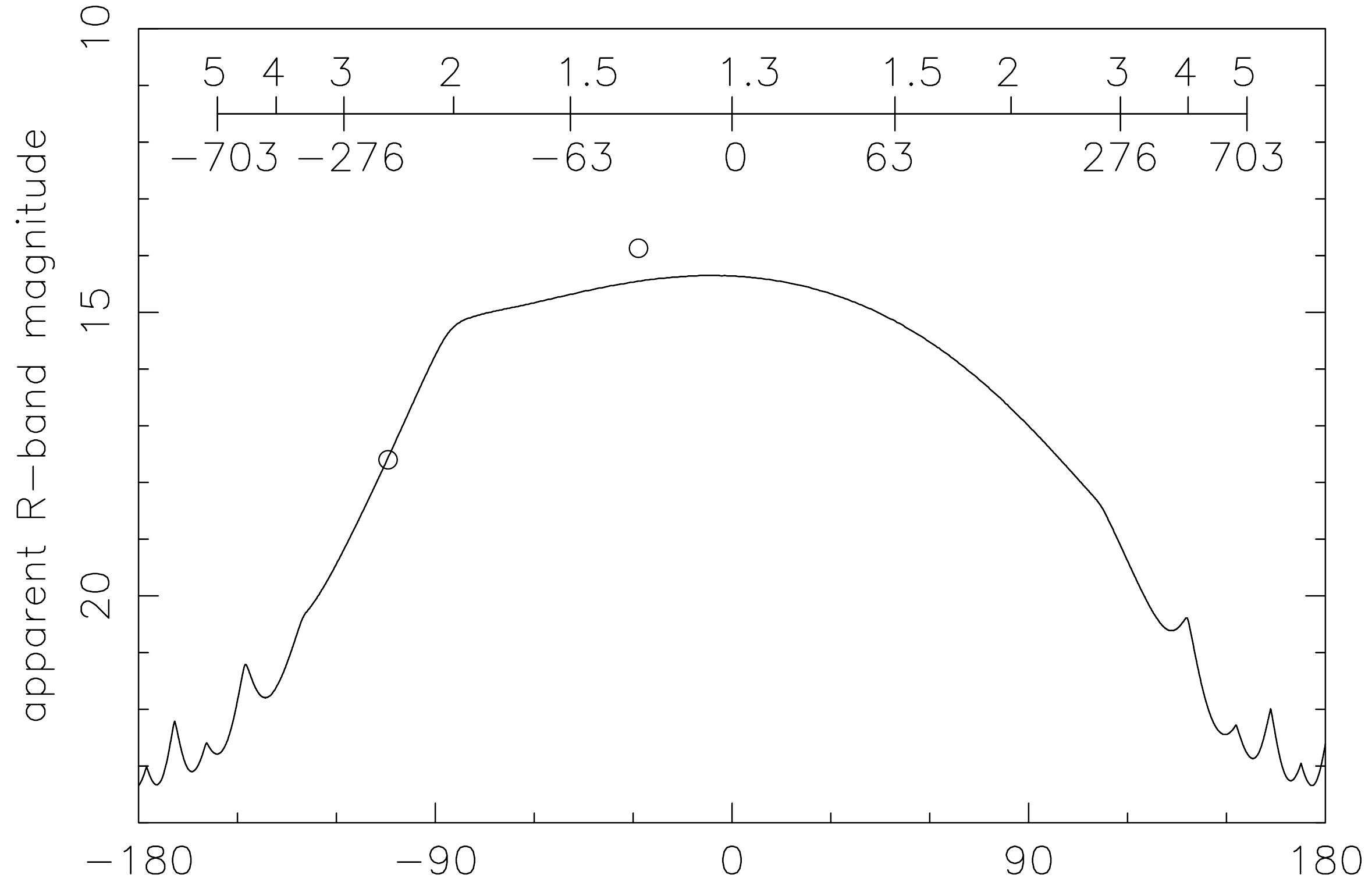} 
   \includegraphics[width=\columnwidth]{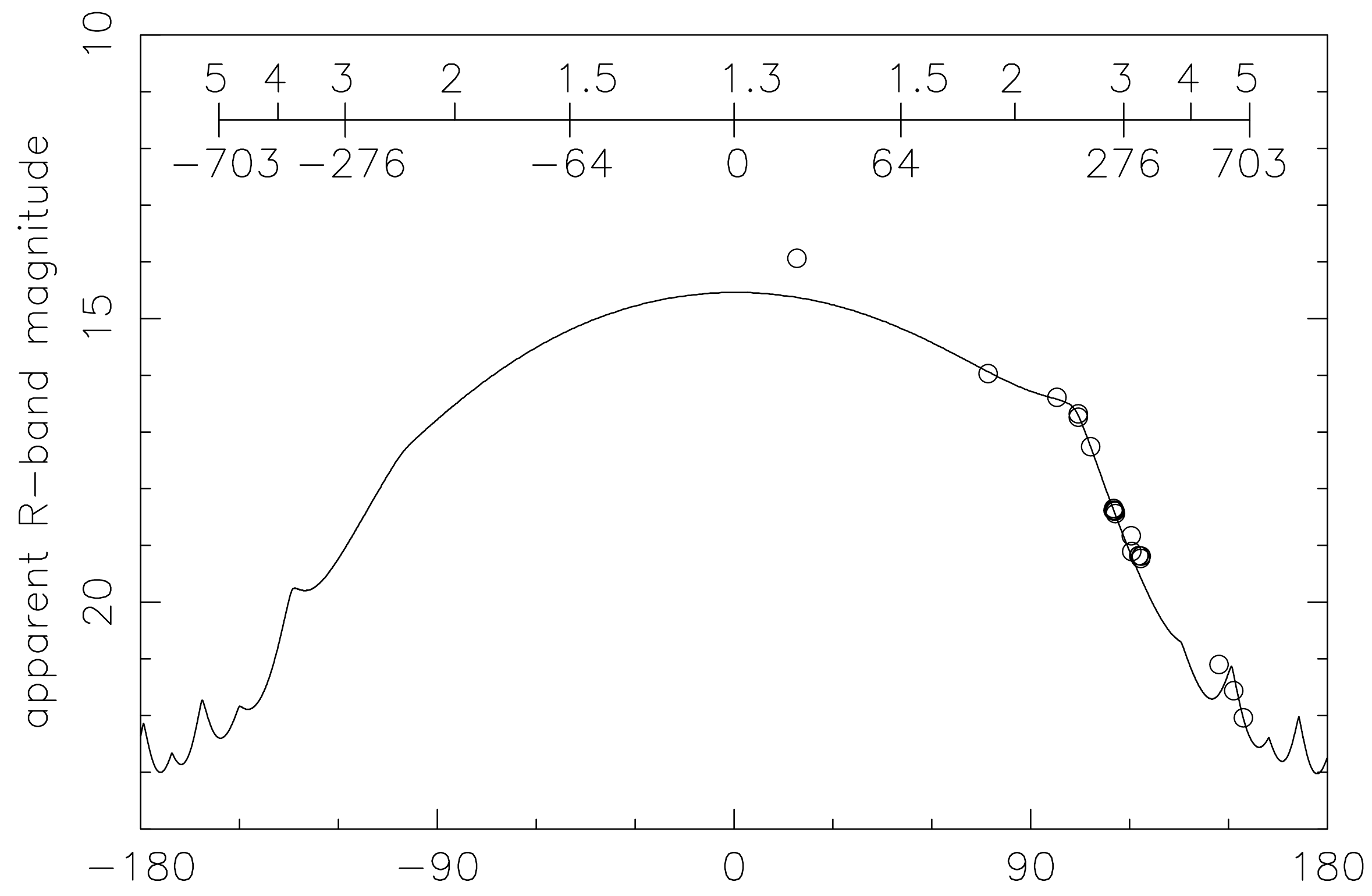} 
   \includegraphics[width=\columnwidth]{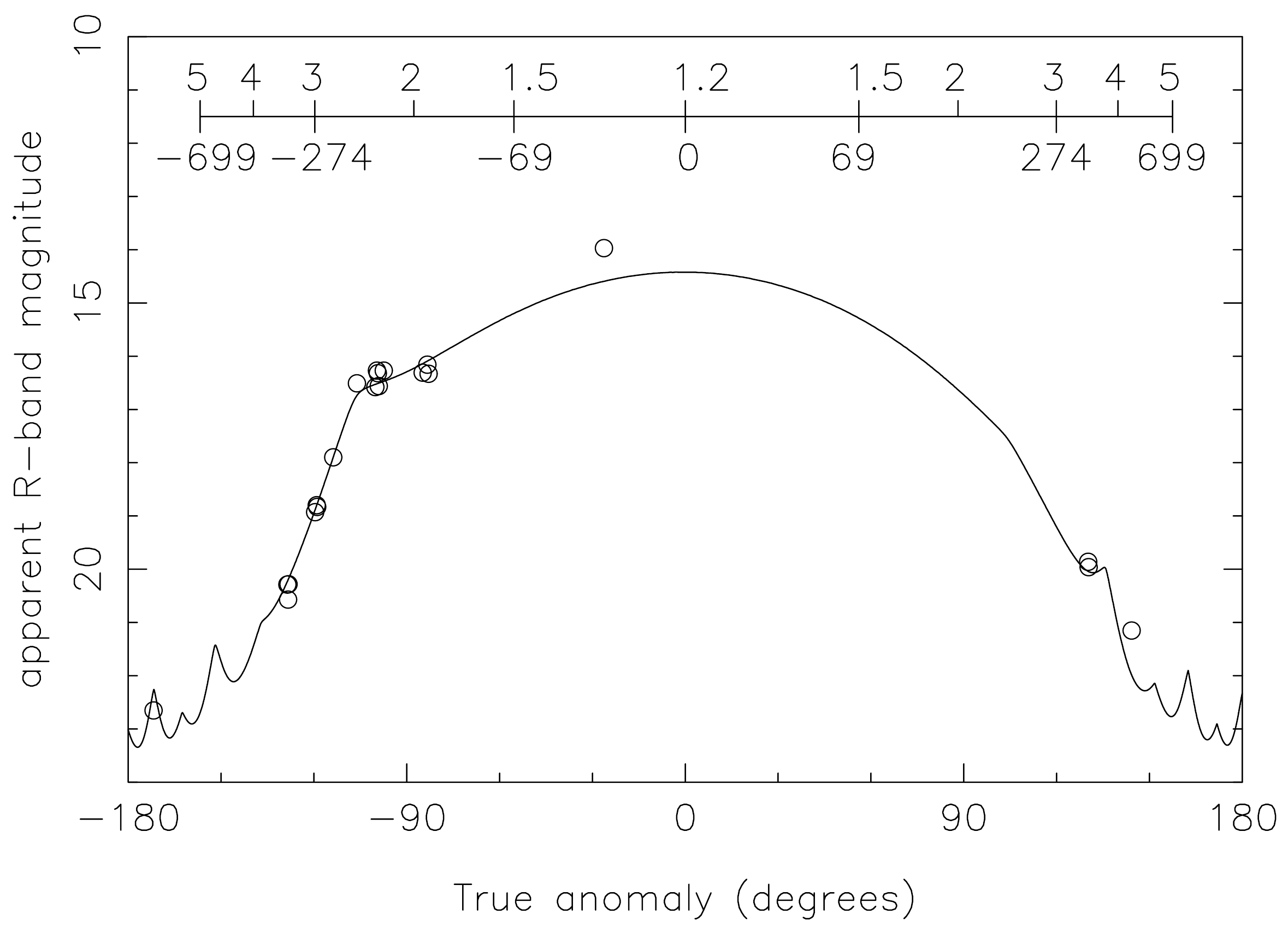} 
   \caption{Model fit to the archival photometry, plotted against true anomaly in degrees. Here the photometry is plotted as apparent magnitudes within $\rho=5\arcsec$ (without correction for observing geometry), as the model includes the geometry (producing `waves' near aphelion due to changing geocentric distance). The three perihelion passages (1996, 2002 and 2009) are plotted separately. The scale across the top of each figure gives the corresponding heliocentric distance (AU) and $\Delta T$ (days). The model reproduces all data well except for that within $\sim30\degr{}$ ($\approx \pm$ 1 month) of perihelion, where there is enhanced activity.}
   \label{fig:model}
\end{figure}

We iterate changing the fractional active surface area until the calculated gas production rates match those observed at specific heliocentric distances.  For CO$_2$ we find a fit consistent with flux estimates from \citet{Bauer12} using WISE observations at 3.32 AU inbound, which gave $Q$(CO$_2$)$ = 5\times10^{25}$ molecules s$^{-1}$, and \citet{Ootsubo12}, who obtained $Q$(CO$_2$)$ = 4.3\times10^{25}$ at 1.84 AU from observations with the Akari satellite.  The models suggested that 0.04\% of the surface area was producing CO$_2$ (inbound) and required an enhancement to 0.09\% post-perihelion.  Matching the photometric light curve brightness required a dominant grain size in the 5-micron range.  The models require that 1.4\% of the surface area is active for water-sublimation, but that the dominant grain size is around 1.5 microns to match the light curve brightness.  Without external constraint on gas production, the light curve brightness can be matched either by increasing the gas production and sublimating area, or by increasing the effective cross section of the scatterers, which can be done by making the average grain size smaller.  The curves shown in fig. \ref{fig:model} are the best fit from the combined CO$_2$+H$_2$O models that are consistent with all of the reported gas production rates, and fix the parameters as shown in table \ref{Kmodel-table}.

The same parameters are used for all three apparitions, showing there is no discernable secular variation in the production.  The three data points close to perihelion are too bright for the model parameters above; the fractional active area has to increase to near 4\% around perihelion, also consistent with the reported water production rates from \citet{Schleicher}.  In fact, the reported water production tapers back down to the values consistent with the model within about 1 month of perihelion.

At large $r$ the production rates predicted by these parameters provide a close match to those given by the fit to $\Delta T$ (equation \ref{Mike-fit}), although they underestimate the perihelion activity. The model gives an enhanced production rates relative to equation \ref{Q-r-eqn} between $r \sim 2$ and 3.5 AU, and lower rates at distances smaller or larger than this, although the results are of the same order of magnitude. At 4.3 AU, the distance we found for the start of `detectable' activity in section \ref{activity_start}, the model gives production rates of $Q$(H$_2$O) = $1.4\times10^{24}$  and  $Q$(CO$_2$) = $7.7\times10^{24}$ molecules s$^{-1}$. Table \ref{kmodel-predict} gives results from the model for various dates and distances, including the dates of relevance to {\sl Rosetta} marked in fig. \ref{fig:predict}.

\begin{table}
\caption{Production rates in 2014/5 from the ice sublimation model.}
\begin{center}
\begin{tabular}{c c c c c c}
\hline
Date & $r$ & \multicolumn{2}{c}{$Q$(CO$_2$)} & \multicolumn{2}{c}{$Q$(H$_2$O)}\\
& (AU) &  kg s$^{-1}$ & molec.~s$^{-1}$ & kg s$^{-1}$ & molec.~s$^{-1}$\\
\hline
20/01/14 & 4.5 & 0.50 & 6.8E+24 & 0.01 & 5.2E+23 \\
20/03/14 & 4.3 & 0.57 & 7.7E+24 & 0.04 & 1.4E+24 \\
23/05/14 & 4.0 & 0.69 & 9.3E+24 & 0.13 & 4.7E+24 \\
20/07/14 & 3.7 & 0.80 & 1.1E+25 & 0.36 & 1.4E+25 \\
20/08/14 & 3.5 & 0.89 & 1.2E+25 & 0.62 & 2.3E+25 \\
11/11/14 & 3.0 & 1.30 & 1.7E+25 & 2.2 & 8.1E+25 \\
21/01/15 & 2.5 & 1.90 & 2.5E+25 & 5.3 & 2.0E+26 \\
28/03/15 & 2.0 & 2.90 & 3.9E+25 & 12 & 4.4E+26 \\
04/06/15 & 1.5 & 5.20 & 7.1E+25 & 27 & 1.0E+27 \\
13/08/15 & 1.2 & 7.70 & 1.0E+26 & 42 & 1.6E+27 \\
\hline
\end{tabular}
\end{center}
\label{kmodel-predict}
\end{table}%

\section{Predictions for 2014/5}

Our observations indicate that activity in 67P starts at large heliocentric distance, at least 4.3 AU inbound. Modelling the full heliocentric lightcurve suggests that there is likely to be a very low level of activity present at even larger distances, which implies that {\sl Rosetta} will find an already active comet when the spacecraft wakes up in January 2014 (at 4.5 AU). We expect the activity to reach a level detectable from Earth (with a large telescope) by March 2014. It will be an interesting test of sensitivities to see whether the remote sensing instruments onboard {\sl Rosetta} are capable of detecting activity before this, although it may not be possible to try, given the expected schedule of recommissioning following deep space hibernation. Once again the comet will appear against a crowded stellar background (low galactic latitude), as seen from Earth, in 2014, necessitating the use of DIA methods to perform photometry from the ground. We expect that the comet will begin to show a noticeable coma in Earth-based images around the same time as the comet's nucleus is resolved by the {\sl OSIRIS} cameras onboard {\sl Rosetta}, in July 2014.

\begin{figure}
   \centering
   \includegraphics[width=\columnwidth]{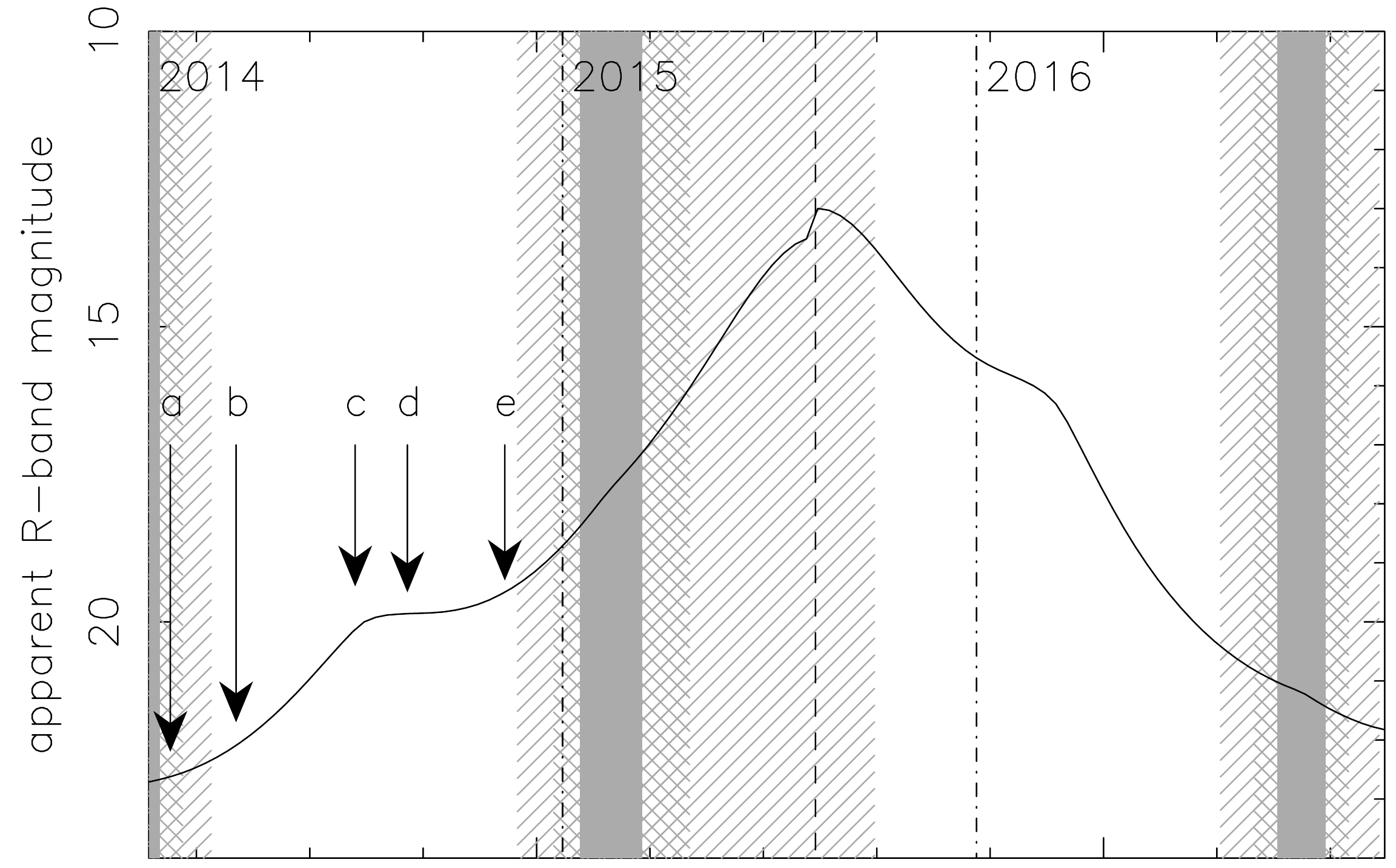} 
   \includegraphics[width=\columnwidth]{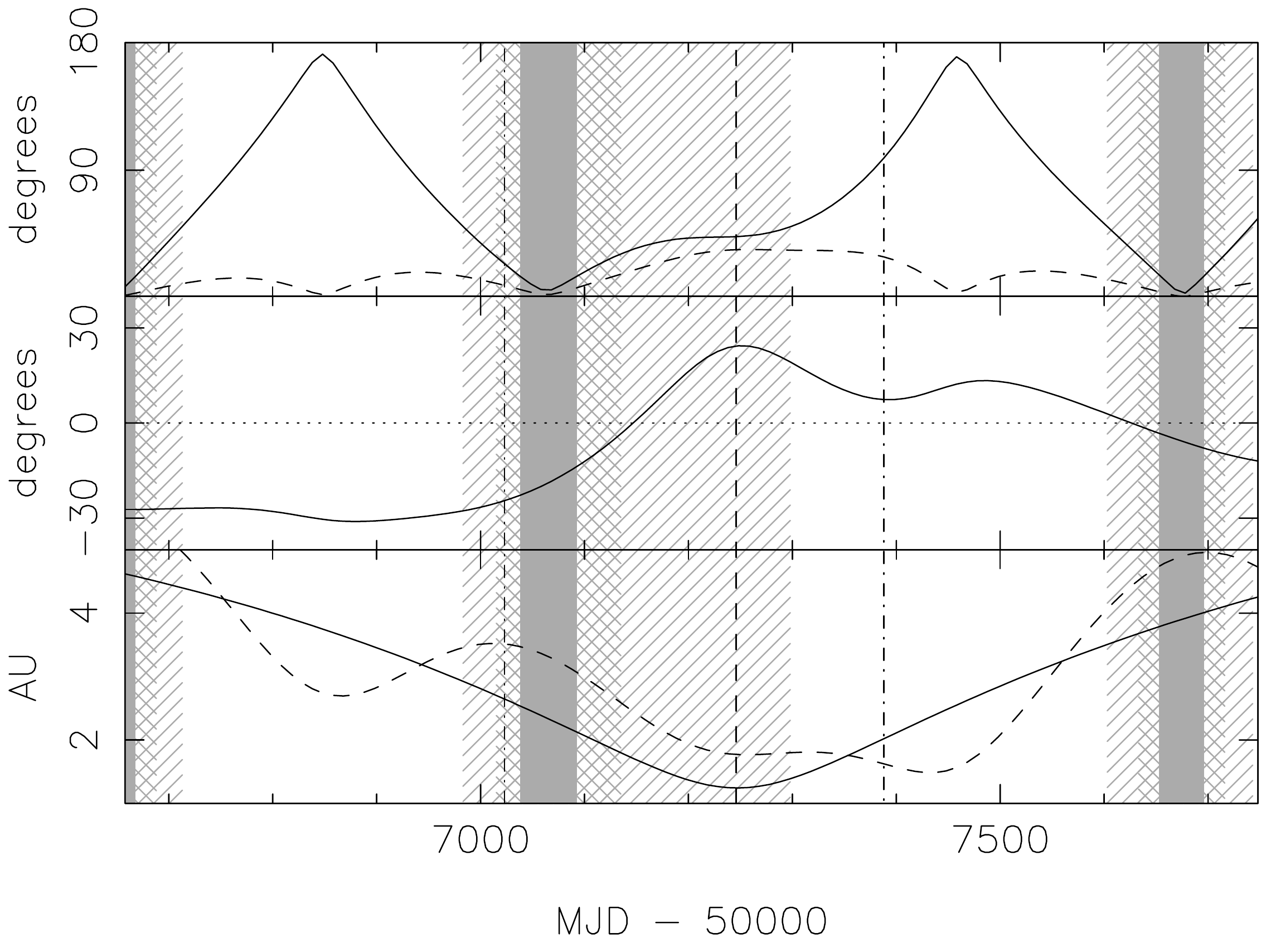} 
   \caption{Predicted apparent $R$-band magnitude of the comet, as measured within an aperture with $\rho=$10,000 km, for 2014-2016. This covers the active phase of the {\sl Rosetta} mission, including any extended mission beyond the end of 2015. Mission milestones in 2014 are marked: (a) Switch on of the spacecraft (January 20th, 2014); (b) Expected beginning of detectable activity, at 4.3 AU (March 2014); (c) The nucleus begins to be resolved by the {\sl OSIRIS} Narrow Angle Camera on {\sl Rosetta} (July 2014); (d) Orbit insertion (August 2014, at 3.5 AU); (e) Lander released (November 2014, at 3 AU). The observability of the comet from Earth is shown by hatched, cross-hatched and solid grey areas marking when the solar elongation is less than 50\degr, 30\degr{} and 15\degr, respectively. Perihelion (in August 2015) is marked by a vertical dashed line. At that time the comet will be 43\degr{} from the Sun, as seen from Earth. The lower 3 panels show various geometric parameters that describe the observability of the comet. Upper panel: Solar elongation (solid line) and phase angle (dashed line); Middle panel: Declination; Lower panel: Heliocentric (solid line) and geocentric (dashed line) distances.}
   \label{fig:predict}
\end{figure}

In fig.~\ref{fig:predict} we show the expected brightness of the comet, as seen from Earth, based on the fit to our heliocentric lightcurve and the observing geometry around the next perihelion. The repeatability of the comet's activity in the last three perihelion passages suggests that we can be reasonably confident about this prediction. We include 2016, to cover the period when the comet is most easily observed from Earth, and the potential extended mission for {\sl Rosetta} beyond its current 2015-12-31 end date. This plot shows that 67P will reach perihelion while at low ($\sim 45\degr$) solar elongation, and consequently be difficult to observe with large professional telescopes. The expected magnitude (peaking at $m_R \approx 13$) is, however, sufficient that total brightness estimates will be possible using smaller telescopes. Production rate measurements, requiring spectra or narrow-band photometry, will be more challenging, but are of great interest. {\sl Rosetta} will provide {\it in situ} measurements of gas abundance in localised areas of the inner coma, but not the wider view needed to link these measurements to ground-based observations of other comets; measurements of the overall production of the comet can be used to make this link. Furthermore, a very complete heliocentric lightcurve through 2015 will allow comparison between the dust production rate of the comet, as measured for the whole body, with the dust flux in the inner coma and the changes in the nucleus measured by {\sl Rosetta}'s instruments. As we have shown, careful consistent analysis and good calibration can result in a very clean lightcurve; if combined with a large campaign to give almost constant coverage then such an approach will reveal any subtle changes (e.g. small outbursts) that can be correlated with events seen by the spacecraft.

\section{Conclusions}

We make use of an advanced difference image analysis package to remove the stellar background in exceptionally crowded fields, to reveal the comet in previously unusable data. This data covers the critical period around $R_{\rm h} \approx 4$ AU when activity is expected to start. 
In addition, we have located archival images of the comet throughout its orbit, and processed them in a consistent manner to produce a reliable heliocentric lightcurve. We find:
   \begin{enumerate}
      \item Detectable activity (using the world's best current telescopes) starts as far from the Sun as 4.3 AU, based on excess flux in photometric measurements, while the comet morphology shows visible activity by 3.4 AU.
      \item The comet's morphology and surface brightness profile exhibit significant variations around the orbit, but the heliocentric light curve is very smooth, with a peak shortly after perihelion.
      \item The dust flux can be described by single power law fits, with only slight differences in slope around perihelion. We find $Af\rho \propto r^{-3.2}$ pre-perihelion and $\propto r^{-3.4}$ post-perihelion.
      \item These slopes are used to predict the dust flux around the next perihelion, with a peak $R$-band magnitude of $\sim 13$ expected in August 2015, as seen from Earth and measured within a $\rho=$ 10,000 km aperture, although the comet will be at relatively low solar elongation ($\sim 45\degr$) at the time. 
      \item By comparing our $Af\rho$ fits with previously published measurements of the gas production rate, we find that the average dust-to-gas ratio for the comet can be expressed as log($Af\rho$/$Q$(H$_2$O)) = $-24.94\pm0.22$ cm s molecule$^{-1}$, for measurements within 1.9 AU of the Sun. A trend of increasing dust-to-gas with increasing distance allows us to find a very approximate $r$ dependence for water production,  $Q$(H$_2$O) $\propto r^{-5.9}$, although we caution that this is unlikely to match reality at larger distances.
      \item A physical model based on sublimation from the nucleus is used to fit the observed brightness of the comet, and suggests that 1.4\% of the surface is active (sublimating water), while 0.04-0.09\% of the surface is sublimating CO$_2$. There is a peak around perihelion requiring an increase of the active area to $\sim4\%$ of the surface.
   \end{enumerate}

\begin{acknowledgements}
We made extensive use of the ESO archive, and wish to thank the staff at ESO Headquarters in Garching who maintain this facility. We are also grateful to all of the original observers and observatory staff who took the data.
We thank Laurie Urban for locating the University of Hawaii 2.2m data on 67P, and Samuel Duddy for providing the rotational phase corrections based on the \citet{Lowry12} model.
We are grateful to Mike A'Hearn, Olivier Hainaut and Stephen Lowry for helpful suggestions. 
The research leading to these results has received funding from the European Union Seventh Framework Programme (FP7/2007-2013) under grant agreement no. 268421. CS also thanks ESO's visiting scientist programme for financial support during a visit to Garching, during which some of this work was carried out.
This research was supported in part by NASA grants NNX13A151G and NNA09DA77A.
We have used the NASA ADS system, including the Dexter graph reading software, and thank the developers and maintainers  for their efforts.
This research used the facilities of the Canadian Astronomy Data Centre operated by the National Research Council of Canada with the support of the Canadian Space Agency. 
Based in part on data collected at Kiso observatory (University of Tokyo) and obtained from the SMOKA, which is operated by the Astronomy Data Center, National Astronomical Observatory of Japan.
This research was made possible through the use of the AAVSO Photometric All-Sky Survey (APASS), funded by the Robert Martin Ayers Sciences Fund.
We made use of an image obtained by the SDSS, and calibrated other data based on their star catalogue. Funding for SDSS-III has been provided by the Alfred P. Sloan Foundation, the Participating Institutions, the National Science Foundation, and the U.S. Department of Energy Office of Science. The SDSS-III web site is http://www.sdss3.org/.
SDSS-III is managed by the Astrophysical Research Consortium for the Participating Institutions of the SDSS-III Collaboration including the University of Arizona, the Brazilian Participation Group, Brookhaven National Laboratory, University of Cambridge, Carnegie Mellon University, University of Florida, the French Participation Group, the German Participation Group, Harvard University, the Instituto de Astrofisica de Canarias, the Michigan State/Notre Dame/JINA Participation Group, Johns Hopkins University, Lawrence Berkeley National Laboratory, Max Planck Institute for Astrophysics, Max Planck Institute for Extraterrestrial Physics, New Mexico State University, New York University, Ohio State University, Pennsylvania State University, University of Portsmouth, Princeton University, the Spanish Participation Group, University of Tokyo, University of Utah, Vanderbilt University, University of Virginia, University of Washington, and Yale University. 

\end{acknowledgements}

\bibliographystyle{aa} % style aa 
\bibliography{comets}  % use bibtex for automatic references section

\end{document}